\documentclass[pdflatex,sn-basic]{sn-jnl}

\jyear{2022}%

\theoremstyle{thmstyleone}%
\theoremstyle{thmstyletwo}%

\theoremstyle{thmstylethree}%

\usepackage{soul}
\begin{document}

\title[Effect of Swarm Density on Collective Tracking Performance]{Effect of Swarm Density on Collective Tracking Performance}


\author*[1,2]{\fnm{Hian Lee} \sur{Kwa}}\email{hianlee\_kwa@mymail.sutd.edu.sg}

\author[3]{\fnm{Julien} \sur{Philippot}}

\author*[3]{\fnm{Roland} \sur{Bouffanais}}\email{roland.bouffanais@uottawa.ca}

\affil[1]{\orgdiv{Engineering Product Development}, \orgname{Singapore University of Technology and Design}, \orgaddress{\city{Singapore}, \country{Singapore}}}

\affil[2]{\orgdiv{Thales Research \& Technology}, \orgname{Thales Solutions Asia}, \orgaddress{\city{Singapore}, \country{Singapore}}}

\affil[3]{\orgdiv{Department of Mechanical Engineering}, \orgname{University of Ottawa}, \orgaddress{\city{Ottawa}, \state{ON}, \country{Canada}}}


\abstract{
%
How does the size of a swarm affect its collective action? Despite being arguably a key parameter, no systematic and satisfactory guiding principles exist to select the number of units required for a given task and environment. Even when limited by practical considerations, system designers should endeavor to identify what a reasonable swarm size should be. 
%
Here, we show that this fundamental question is closely linked to that of selecting an appropriate swarm density. Our analysis of the influence of density on the collective performance of a target tracking task reveals different `phases' corresponding to markedly distinct group dynamics. We identify a `transition' phase, in which a complex emergent collective response arises. Interestingly, the collective dynamics within this transition phase exhibit a clear trade-off between exploratory actions and exploitative ones. We show that at any density, the exploration-exploitation balance can be adjusted to maximize the system's performance through various means, such as by changing the level of connectivity between agents. 
%
%
While the density is the primary factor to be considered, it should not be the sole one to be accounted for when sizing the system. Due to the inherent finite-size effects present in physical systems, we establish that the number of constituents primarily affects system-level properties such as exploitation in the transition phase.
%
%
These results illustrate that instead of learning and optimizing a swarm's behavior for a specific set of task parameters, further work should instead concentrate on learning to be adaptive, thereby endowing the swarm with the highly desirable feature of being able to operate effectively over a wide range of circumstances. 
}




\maketitle
\section{Introduction}

Decentralized multi-robot systems (MRS) have recently become a highly active field of study owing to their unique ability to tackle a number of critical and challenging problems where a task has to be accomplished in a highly dynamic environment. When facing rapidly evolving conditions, the effectiveness of an MRS critically hinges on its ability to change behavior owing to its flexibility and/or adaptivity~\cite{Dorigo2021}. This includes a wide range of problems such as target tracking~\citep{Coquet2019, Coquet2021}, area protection~\citep{Strickland2018, Shishika2019}, dynamic area monitoring~\citep{Vallegra2018, Zoss2018}, area mapping~\citep{Kit2019,Crosscombe2021, Liu2021}, and environment classification~\citep{Ebert2018, Ebert2020}. Beyond flexibility and adaptivity, the effectiveness of an MRS has also been shown to be related to robustness and scalability~\citep{Bouffanais2016, Hamann2018swarmrobotics, Dorigo2021}. These key attributes stem from the lack of a central controller that dictates the actions of the individual agents and have made MRS very attractive when considering operations in some of the most challenging circumstances involving expansive, unstructured, and dynamic environments. The decentralized nature of such systems divides up the overall control of the MRS and gives individual robots the ability to make decisions based on their local environment. This allows them to: (1) quickly react to changes in their surroundings (flexibility), (2) continue operations despite agent failures given the lack of a central point of control (robustness), and (3) eliminate potential processing bottlenecks (scalability). In addition, an MRS also has the ability to learn or change its behavior to deal with new circumstances (adaptivity). 

One key design consideration and a critical question that practitioners must contend with is selecting the number of robotic units required to achieve a desired performance for a given task and environment~\citep{Hamann2018swarmrobotics}. Indeed, it has been shown that in both biological and engineered multi-agent systems, effective emergent collective actions require a critical number of swarming agents. Specifically, \cite{Schranz2021} stated that ``the advantages of swarm intelligence algorithms can only be exploited if a critical mass of swarm members is reached." However, a system's collective behavior is not the only consideration when determining a swarm's size. For a given task, several other aspects also influence the selection of the number of swarming agents, such as: (1) system scalability, (2) technical capabilities of the individual agents (e.g., communications range, sensor range, maneuverability, maximum speed, etc.), and (3) financial or logistical constraints (i.e., the number of robots that can be built, stored, and operated given the available resources)~\citep{Schroeder2019}. It can therefore be said that selecting the number of robotic units is far from being a trivial task for system designers.

In the literature, MRS come in various shapes and sizes, with their sizes covering three orders of magnitude. These range from the 1024-unit Kilobot swarm by \cite{rubenstein2014kilobot}, to the 150 unit subCULTron system by \cite{Thenius2016}, to the hunter drone swarm by \cite{DeSouza2022} that can operate with less than 10 units. It should be noted that there is no consensus on what constitutes a robot swarm in terms of the number of units~\citep{Hamann2018swarmrobotics}. However, in the vast majority of works, it is often unclear how exactly this quantity is selected. Would a half-sized Kilobot swarm perform vastly differently from the full-sized Kilobot MRS~\citep{Rausch2019}? At the other end of the spectrum, the same question can be posed to system designers who utilize relatively small systems, e.g., with 10 or less units. While it is often argued that such a small system size would be sufficient for the task at hand~\citep{Zoss2018, Kit2019, Kwa2021}, one is often left wondering if the limit in the actual number is not related to other factors, such as the logistical, financial, or technical challenges discussed above.

The size of the system is inextricably linked to the density of agents. Many works hint at the existence of a minimum critical density required to yield an effective self-organization of the swarm. For example, the agents used by the Kilobot swarm are only able to communicate with each other at a maximum distance of 10~cm~\citep{rubenstein2014kilobot}. Similarly, it was determined by \cite{Hornischer2020} that for a robot to function as intended in their system, a minimum of 3 other robots had to be located within an agent's communication range. Beyond the constraints in communications or information exchange, a scaling law has been used to quantify the effectiveness of the BoB system when performing a dynamic area monitoring task with 50 buoys for a pre-determined surface area~\citep{Zoss2018}.

Fundamentally, the question of the selection of an MRS size---or equivalently its agents density---for a given task has never been systematically and satisfactorily answered, if at all. As already mentioned, the importance of a minimum density has long been acknowledged to ensure the emergence of a desired collective behavior in simplistic physics models. The self-propelled particles (SPP) in Vicsek's model is a paradigmatic example of that~\citep{Vicsek1995}. Irrespective of the amount of noise present, it was shown that the agent density is key in ensuring the self-ordering process. Specifically, at low density, the swarm of SPP exhibits a disordered state that undergoes a phase transition leading to an ordered phase---with all SPP being practically aligned---when increasing the density of agents. Beyond the simplistic case of SPP seeking to align their direction of travel, \cite{Hamann2012} considered the effect of swarm size and density---the surface area being kept constant---on system-level performance for several collective behaviors and decision-making protocols. This revealed superlinear performance increases when the system was enlarged, followed by a decrease in performance once the swarm exceeded a certain size. This degradation in performance was attributed to the increasing effect of interference on the system's performance, resulting from the increased amount of collision avoidance behavior carried out by the agents. 

Besides collective motion, the problem of a system's exploration-exploitation balance, which is an important factor in determining the overall swarm performance, also needs to be considered. In various computational optimization tasks, such as virtual agents carrying out particle swarm optimization, and in the operation of MRS, larger swarms have been shown to exhibit higher levels of exploratory activity~\citep{Shishika2019, Piotrowski2020}. However, this large amount of exploration needs to be paired with an adequate amount of exploitation to make use of the information gathered during the exploration process.

In this paper, we study the effects of density on the effectiveness in target tracking by an MRS: a highly dynamic collective task. Given the importance of a swarm's exploration and exploitation balance, we also examine the influence of swarm density on this balance, and show how it ultimately affects the overall system performance. It can be argued that in past studies, some of the tasks considered were too undemanding and could be accomplished using independently operating units, without the need for self-organization or emergent collective actions. However, we believe that it is only through dealing with such challenging problems that a system can demonstrate the full power of swarm intelligence. Therefore, this study is conducted using a fast-moving target (i.e., one that moves faster than any of the swarm's component agents) that does not emit a signal gradient field, making it akin to a visual search task where a target needs to be seen to have its presence confirmed~\citep{Esterle2020}. Doing so eliminates the possibility of using simplistic methods such as gradient descent to track the target, thus forcing the system to self-organize and behave as a collective to track the target. It should be noted that while there are stochastic methods that attempt to control agent density, such as those using Markov processes~\citep{Acikmese2012, Elamvazhuthi2019, Biswal2021}, in this paper, we focus mainly on strategies employing deterministic approaches.

Our investigation reveals three density `phases' that characterize the tracking ability of an MRS, namely: (1) a low density phase where the system is simply unable to coordinate the agents' behavior to carry out the tracking task, (2) a transition phase where the performance of the system rises very quickly with increasing swarm density, and (3) a high density phase where the system is able to continuously track the target without any interruptions. These phases occur regardless of the number of agents within the system. We explain these trends by the existence of a system-level exploration-exploitation trade-off within the transition phase, with clearly identifiable exploration-limited and exploitation-limited regimes. Interestingly, the level of connectivity among agents can be used to skew the swarm dynamics towards either exploration or exploitation. Specifically, increasing (resp. decreasing) connectivity boosts the tracking performance in the low density exploitation-limited (resp. high density exploration-limited) regime. Nonetheless, the system size still plays a role in determining the ability of the MRS to carry out exploration and exploitation. Lastly, we discuss the significant ramifications that these results have for the development of effective collective dynamics through machine learning and automated design.

\section{Related Work}
\subsection{System Size and Exploration}
The number of units employed within a multi-agent system is one of the basic parameters that a system designer needs to select. In MRS or virtual multi-agent systems (MAS), studies on a system's size usually occurs within an environment of a fixed size. As such, adding agents into a system equates is equivalent to an increase in agent density, and conversely removing agents results in a system with a lower swarm density. In the field of optimization, the number of agents, sometimes known as candidate solutions, employed is synonymous with the amount of exploration carried out by a system~\citep{VanDenBergh2001, Piotrowski2020}. As can easily be understood, using a larger number of agents allows the system to more thoroughly explore the search space, thereby reducing the chances of the optimization algorithm to settle in local optima. Similar observations were made when studying bee colonies in the wild where it was observed that larger groups were able to employ more scouts, thereby increasing the amount of exploration carried out and the accuracy of the colony's final consensus~\citep{Schaerf2013}. These results were also observed in engineered MRS, with system performances tending to increase with larger system sizes~\citep{Kit2019, Prasetyo2019, Shishika2019, Jurt2022}. Similar to the virtual MAS used in optimization and those found in Nature, these increases in performance are attributed to the increased exploration carried out by the larger systems.

\subsection{Marginal Utility}
Given that larger MRS tend to yield systems with better performance, some studies have focused on the marginal utility gain when adding agents. These works focus on the marginal utility gain under the context of resource foraging. One of the earliest works on this topic was done by \cite{Lerman2001}, who demonstrated that while additional robots may indeed improve the foraging performance of the group as a whole, the amount of resources gathered by each robot, i.e., their individual efficiency, decreases. This is due to the increased level of physical interference in larger systems caused by the robots colliding with each other. Similar results were also demonstrated by \cite{Kit2019} in a collective area mapping task and by \cite{Sung2018, Sung2020} in a target tracking scenario. However, in target tracking, unlike resource foraging and area mapping, the swarm performance no longer increases after the MRS reaches a certain size. This is because at this point, the system is able to track the target regardless of how it moves. As such, the addition of units into the MRS provides no extra performance gain for the system. Interestingly, even in MAS optimization, it is often reported that increasing the number of agents beyond a certain point yields only marginal benefits to the system's accuracy and notably lengthens computational times~\citep{roeva2015population}.

The reducing robot efficiency with increased system size and higher agent densities is known in economics circles as the ``Law of Diminishing Marginal Returns'' and was modeled in MAS by~\cite{Hamann2012}. This was done using an exponential function that accounted for both the benefits of inter-agent cooperation and the detriments caused by the additional interference. Subsequently, \cite{Schroeder2019} combined this multi-agent performance strategy with a simple cost model to maximize the performance of their foraging MRS while minimizing the overall system cost by finding the ideal number of robots to be constructed for their system.

The diminished marginal utility caused by the addition of more agents may be further aggravated if common resources are to be shared and exploited. In a foraging task, \cite{Hecker2015} concluded that when agents had to deposit their resources at a central location, individual robot efficiency decreased further in large swarms due to robots being required to travel longer distances to obtain resources. In \cite{Rosenfeld2006}, the authors showed that the over-exploitation of common resources results in a higher level of physical interference between agents. In turn, this increased interference may result in negative marginal utility, i.e., the performance of the system decreases when more agents are included within the MRS. This occurs despite the higher amount of exploration taking place due to the increased swarm density and is because the advantages provided by the additional exploration is overshadowed by the exponentially larger level of interference encountered by the individual agents. This result was also demonstrated by \cite{Hamann2018} in a stick-pulling task, allowing the author to demonstrate three separate phases when adjusting a swarm's size: (1) a phase where agents do not collaborate enough and do not make full use of the available resources, (2) a phase where agents share too much information, resulting in the depletion of shared resources, and (3) an intermediate phase where the resources are neither underutilized nor overutilized, resulting in the optimal performance of the system.

\subsection{Motility-Induced Phase Separation}
The increased amount of physical interference between agents is more apparent in systems tasked with clustering around certain points (e.g., target tracking and consensus tasks). As previously mentioned, high levels of clustering results in higher levels of physical interference in areas of high local swarm density. This in turn results in a positive feedback loop, causing more agents to aggregate in preexisting areas of high local agent densities in a phenomenon known as Motility-Induced Phase Separation (MIPS)~\citep{Cates2015}. The buildup of agents in specific areas results in excessive amounts of exploitation, thus degrading the system's performance. To combat such excessive levels of exploitation caused by MIPS, several strategies have been developed to take into account the local swarm density around individual agents. When using an MRS to identify areas of high light intensity, modifications were made to the original BEECLUST algorithm, allowing robots to move off from its waiting position should it determine that it is located within an area with high agent density. This subsequently prevented excessive flocking of agents and encouraged a higher level of exploration, thereby making the system more responsive to dynamic environment changes~\citep{wahby2019collective}. \cite{Li2017} developed a strategy to control the positional distribution of robots in an MRS based on local swarm density and the strength of a signal field. To carry out a foraging task, \cite{pang2019swarm} varied the step size taken by each robot based on its local swarm density estimation. In target search and track tasks, many strategies have incorporated inter-agent repulsion to prevent the excessive flocking of robots at a single location, thereby promoting area exploration~\citep{Zhang2019, Dadgar2020, Kwa2020a}.

\section{Strategy and Simulation}
\subsection{Search and Track Strategy}\label{sec:search-and-track-strategy}
The strategy used to facilitate cooperative tracking of fast-moving targets was previously presented in \citep{Kwa2021, Kwa2022c}. This strategy consists of two main components: (1) agent aggregation augmented with a short-term memory and, (2) inter-agent adaptive repulsion, from which the system's adaptivity is derived. These two components are responsible for directing system exploitation of information and environment exploration respectively and is done by generating two velocity vectors at each time-step that are combined to give a final agent velocity vector:
\begin{equation}
    \bv_i[t] = \bv_{i,\text{att}}[t] + \bv_{i,\text{rep}}[t],
    \label{eqn:movement}
\end{equation}
where $\bv_{i,\text{att}}[t]$ and $\bv_{i,\text{rep}}[t]$ are the velocity vectors generated by the attractive component and the repulsion component respectively. The final velocity vector, $\bv_i[t]$, is then scaled by $v_{\text{max}}$, the maximum speed of the agent. The overall exploration and exploitation balance, or exploration-exploitation dynamics (EED), of the system is controlled via the adjustment of the degree of the interconnecting topological $k$-nearest neighbor communications network. This adjustment and its effect on the EED will be further discussed in the next section.

The overall strategy employed in the system is summarized in Algorithm~\ref{alg:strategy}. For further details regarding the strategy's components, the reader is referred to \cite{Kwa2021}.

\begin{algorithm}
    \caption{: Dynamic $k$-Nearest Network Search and Tracking Strategy}
    \label{alg:strategy}
    \begin{algorithmic}[1]
        \State Set $t = 0$, $k \in [2, N-1]$, $\omega=1$, and $ c=0.5$
        \While{System active}
            \For{All agents}
                \State Set point of attraction
                \State Calculate $\bv_{i,\text{att}}[t]$
                \State Calculate $\bv_{i,\text{rep}}[t]$
                \State $\bv_i[t] \gets \bv_{\text{att},i}[t] + \bv_{\text{rep},i}[t]$ // Apply Eq.~\eqref{eqn:movement}
                \State $\bv_i[t] \gets (v_{\text{max}}/v_i[t]) \cdot \bv_i[t]$ // Ensure magnitude of velocity vector does not exceed maximum speed
                \State $\bx_i[t+1] \gets \bx_i[t] + \bv_i[t]$ // Update agent position
            \EndFor
            \State $t \gets t+1$
        \EndWhile
    \end{algorithmic}
\end{algorithm}

\subsection{Network Connectivity} 
\label{sec:network}
A key ingredient to any MRS is the communications network used by individual robots used to share information around the system~\citep{Sekunda2016}. While most research on MRS have either assumed agents with unlimited communications range~\citep{Coquet2021, Rossides2021}, or included all other robots within communications range~\citep{Dadgar2017, Jensen2018, Dadgar2020}, it has been shown that tuning the network connectivity of a system can have considerable effects on a swarm's collective dynamics. \cite{Mateo2017} demonstrated that increased levels of communication resulted in reduced collective response to a simulated predator attack. Furthermore, \cite{Mateo2019, Horsevad2022} also showed that while systems with higher levels of communication were more adept at collective mimicking of a slow-moving leader, lower levels of connectivity were required to more accurately replicate the movements of a fast-moving driving agent. Similar increases in performance were also observed in a scenario where agents were required to accurately characterize a static and noisy environment~\citep{Crosscombe2021, Liu2021, Kwa2022b}.

Building off these conclusions, we previously showed that there exists an optimal level of connectivity, $k^*$, of a $k$-nearest neighbor network at which the performance of a swarm is maximized while tracking a fast-moving target~\citep{Kwa2020a, Kwa2020b, Kwa2021, Kwa2022c}. At this optimum, an MRS is able to attain the ideal balance between the level of exploratory and exploitative activities that its component agents carry out. This value of $k^*$ changes depending on the conditions the MRS is operating in. It should be noted that a neighborhood is understood in the network sense; an agent $i$ has as many neighbors as its degree $k$. Given that time-varying network topologies are considered, the neighborhoods of each individual agent are dynamic and change over time.

\subsection{Problem Statement}
In this work, a set of $N$ tracking agents ${A=\{a_1, a_2, \ldots, a_N\}}$ and a single target, $o$, move within a bounded two-dimensional square search-space of dimensions $L \times L$ devoid of any obstacles, where $L \in [10^{0.6}, 10^{2.65}]$. The agents' and target's positions are denoted by ${\bx_i = (x_i, y_i)}$, and maximum velocities of $\bv_{a, \text{max}}$ and $\bv_{o, \text{max}}$ respectively. It should be emphasized that in this work, we consider a target that moves faster than its pursuing agents. As such, ${\bv_{a, \text{max}} < \bv_{o, \text{max}}}$. In this work, the maximum velocities are set as follows: ${\bv_{a, \text{max}}} = 0.1$ and ${\bv_{o, \text{max}} = 0.15}$.

The target is modeled using a disc-shaped binary objective function with a fixed radius of ${r = 1}$. A target is considered to be tracked if an agent lies within its radius. Formally: 
\begin{equation}
\label{eqn:coverage}
    \text{cov}(o, t)=
    \begin{cases}
        1, & \exists i \in A \text{ s.t. } \|\bx_i - \bx_o\| \leq r, \\
        0, & \text{otherwise.}
    \end{cases}
\end{equation}
The modeling of the target as a binary objective function, as done here, makes the problem more challenging and more similar to a visual search task where a target needs to be seen to have its presence confirmed~\citep{Honig2016, Esterle2020}. This is in contrast to tracking a target from the intensity of an emitted signal (e.g., radio signal strength, chemical plume, etc.) in which various gradient-descent methods can be used. While such gradient-descent methods form one of the most widely used classes of techniques tracking targets, such techniques become completely ineffective when dealing with such binary objective functions. The use of this type of function ensures that such simplistic techniques cannot be used by the system. This conservative approach represents one of the most challenging cases with a near-zero-range sensor tracking a target that is moving faster than the agent themselves. As already mentioned, it can be argued that the full power of swarm intelligence can only be accessed when dealing with such challenging scenarios.

The target is set to move according a non-evasive movement policy. Using this policy, a target travels toward a randomly generated waypoint within the search space. Upon reaching this waypoint, another waypoint is generated, causing the target to change its direction of travel. This process is repeated until the end of the simulation.

The goal of the system is to maximize its tracking performance within the environment, given by the reward function:
\begin{equation}
\label{eqn:cost_fn}
    \Xi = \frac{1}{T} \sum^T_{t=1}\text{cov}(o, t),
\end{equation}
where $T$ it the total time period of interest, set as $T=100,000$. This large value of $T$ was set to ensure the statistical stationarity of the results, i.e., the system that we are testing is ergodic and given that the conditions stay the same, its tracking performance will tend to a constant value when $T \rightarrow \infty$. In the simulations, the agents are tasked with tracking the targets in an environment free from obstacles and are assumed to have perfect information about the target's location once within the target's radius. While carrying out the tracking task, the agents are also assumed to have perfect information about their pose in the environment with respect to a global reference frame.

\subsection{Metrics}
\label{sec:metrics}
In addition to the MRS tracking performance, its collective dynamics can also be studied through the quantification of the EED of the system. As such, the reward function presented in Eq.~\eqref{eqn:cost_fn} is used to quantify how well a system exploits the targets' positional information. 

To further quantify a system's collective dynamics, its \textit{exploration ratio} is also calculated. An agent is considered to be exploring the environment if it has entered the `exploratory' state and $S_{i,\text{exp}}[t] = 1$. In this exploratory state, the agent tries to repel itself away from its neighbors, causing it to explore the environment in the process. In contrast, when an agent detects a target, it enters a `tracking' state, $S_{i,\text{exp}}[t]$ is set to $0$, and the agent attempts to follow the target. Therefore, the overall exploration ratio of the swarm is calculated as follows:
\begin{equation}
    \label{eqn:engagement}
    \Theta = \frac{1}{NT}\sum^T_{t=1}\sum^N_{i=1} S_{i,\text{exp}}[t],
\end{equation}
where $N$ is the total number of agents within the system. With the exploration ratio, a higher value of $\Theta$ indicates a higher proportion of agents spending more time searching for a target, and hence higher levels of exploration. Conversely, at lower $\Theta$, agents attempt to move towards a target, thus spend more time exploiting target information.

The concept of system's \textit{swarm density} or \textit{agent density} is also introduced. It should be noted that these two terms are used interchangeably in this paper. This is calculated as follows:
\begin{equation}
    \label{eqn:dens}
    \rho = N/L^2.
\end{equation}
Moreover, the concept of \textit{local swarm density} is quantified by means of the measure of the average distance from each agent to its 6 nearest neighbors (see Appendix for further details about the choice of this particular value). This allows for the calculation of the local swarm density for each agent. The local density is then averaged for all agents in the swarm across the entire duration of the simulation:
\begin{equation}
    \label{eqn:loc-dens}
    \rho_L = \frac{1}{NT}\sum^T_{t=1}\sum^N_{i=1}\frac{7}{\pi L_{i,t}^2},
\end{equation}
where $L_{i}$ is the average distance from an agent to its six closest neighbors. Using $\rho_L$, one can quantify the level of clustering carried out by a system's agents. This is done by finding the difference between the swarm density and the average local swarm density:
${\Delta\rho = \rho_L - \rho}$, referred to as the density difference.

\section{Results}
\label{sec:results}
\subsection{Effect of Swarm Density on Tracking Performance}
\label{sec:dens_track}

\begin{figure}[hbtp]
    \centering
    \includegraphics[width=\linewidth]{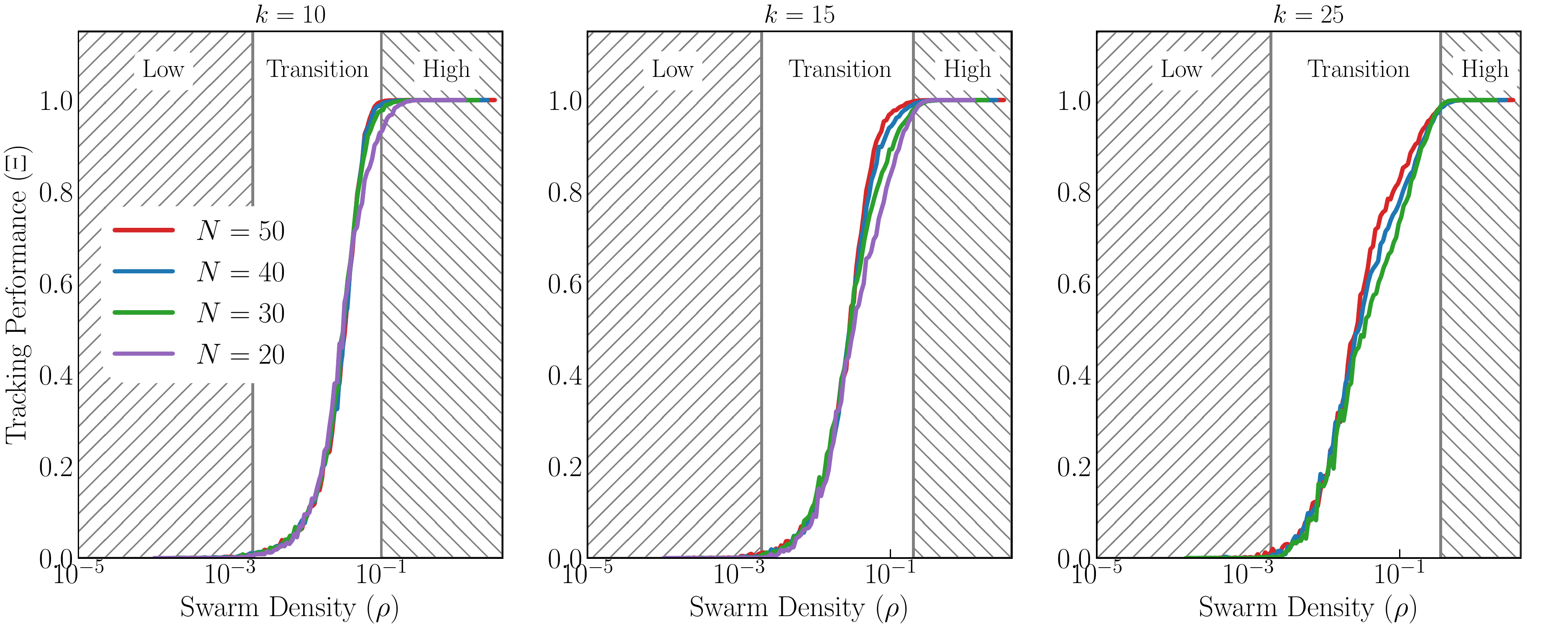}
    \caption{A swarm of $N=20$, 30, 40, and 50 agents with varying levels of connectivity, $k$, and swarm density, $\rho$, tracking a fast-moving non-evasive target traveling at $\bv_{o, \text{max}} = 0.15$. In the low density phase, the system is unable to track the target. In the high density phase, the system able to track the target continuously without interruptions. Between these two phases, there is a transition phase in which the tracking performance of systems rapidly increase with swarm density, $\rho$.} 
    \label{fig:score-size}
\end{figure}

To investigate the effects of swarm density on tracking performance, we first run simulations of different swarm sizes, ranging from $N=20$ to $N=50$, tracking a non-evasive target in different environment sizes. In Fig.~\ref{fig:score-size}, three separate `phases' can be observed. At low density (${\rho \lesssim 10^{-3}}$), the system is effectively unable to track the target. This is due to the large distances between agents, preventing the units from effectively clustering together and tracking the target. The large inter-agent distances also reduce the chances of an agent encountering the target, further aggravating the problem of poor tracking performances. It is worth stressing that a key assumption of our model is that the quality of information exchanges remain unaffected by the large distances between agents---the agents keep the same number of topological neighbors, $k$, and communications are assumed to be perfect. At high density (${\rho \gtrsim 10^{-1}}$), the system is able to carry out near-perfect tracking of the target. At these densities, due to the small inter-agent separation and the lack of open space for the target to move into, the agents are able to constantly keep track of the target.
\begin{figure}[hbtp]
    \centering
    \includegraphics[width=\linewidth]{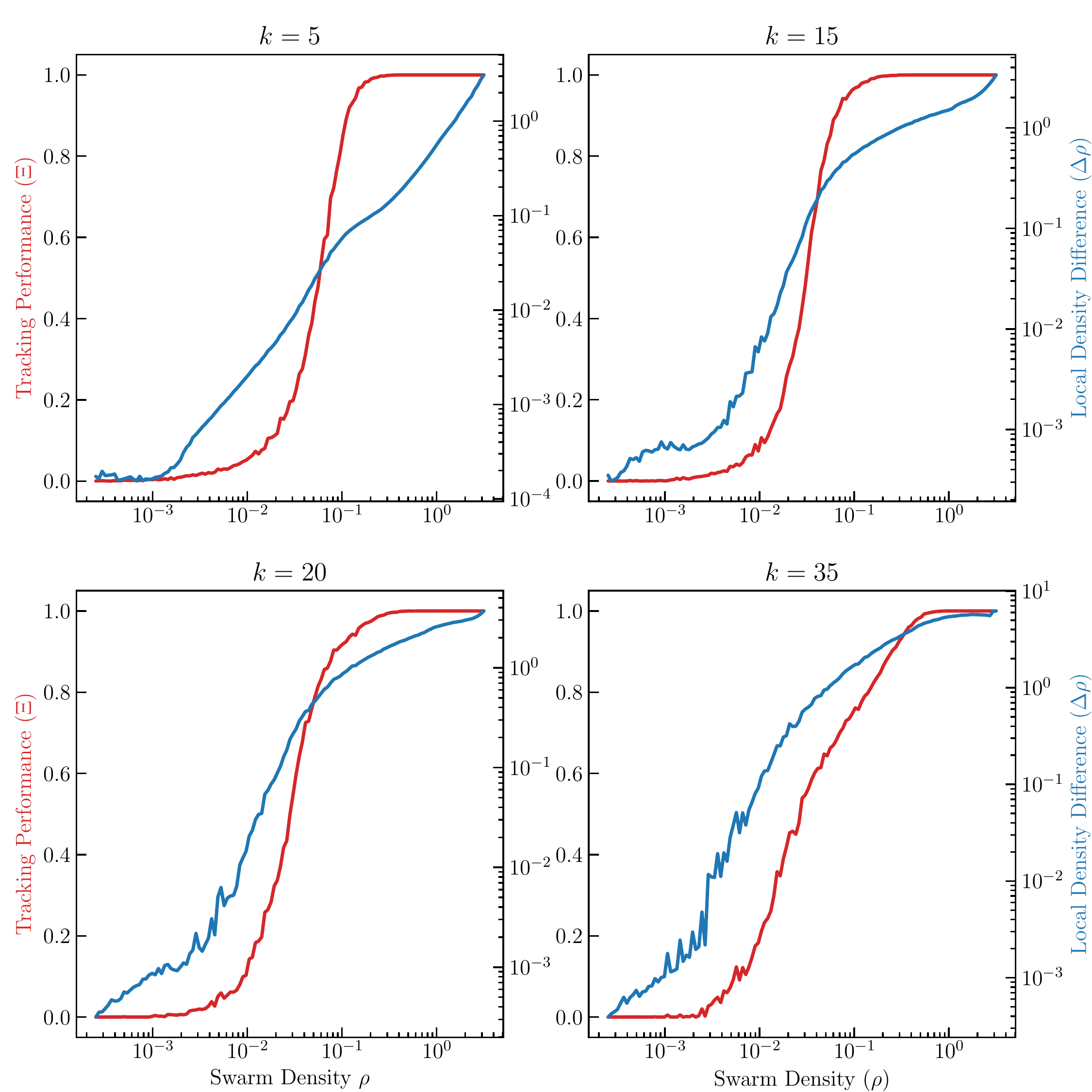}
    \caption{Density difference, $\Delta \rho$, of a swarm comprised of $N=50$ agents with various degrees of connectivity $k$, tracking a fast-moving non-evasive target traveling at $\bv_{o, \text{max}} = 0.15$.} 
    \label{fig:loc-dens_diff}
\end{figure}

Between these two phases sits a `transition' phase in which the tracking performance, $\Xi$, sharply rises with $\rho$. The start of this transition phase can be traced to the ability of agents to effectively cluster around a point of attraction and is important as it allows new agents to takeover the tracking task from the agents that initially encountered the target. This ability to cluster is revealed by the increase in density difference, $\Delta \rho$, (introduced in Sect.~\ref{sec:metrics}) versus $\rho$ (Fig.~\ref{fig:loc-dens_diff}). The density at which the transition phase starts appears to be independent of $N$ (Fig.~\ref{fig:score-size}). The inability of the agents to cluster and make use of target information suggests that at low density, the system's performance is limited by its lack of exploitation. This is because when the level of connectivity, $k$, is constant, the same number of agents are drawn to the target. However, when approaching the high density phase, large swarms (i.e., high values of $N$) outperform smaller ones (i.e., low values of $N$) since the latter tend to dedicate more resources---i.e., a larger proportion of agents---to exploiting target information, thus leaving very few agents to carry out area exploration. Therefore, it can be said that these small systems tend to have their tracking performance limited by a lack of exploration. In summary, low density swarms are exploitation-limited, while high density ones are exploration-limited. The implications of swarm size and its level of connectivity on the systems EED is explored in the subsequent sections.

\subsection{Effect of Swarm Density on Exploration-Exploitation Dynamics}
\label{sec:eed}

To determine the effects of changing a swarm's density on its exploration-exploitation dynamics (EED), we characterize the system's EED by calculating an \textit{exploration ratio}, $\Theta$, which is a system-level metric (introduced in Sect.~\ref{sec:metrics}). By construction, high $\Theta$ values indicate that the swarm is biased towards carrying out exploration. Conversely, low $\Theta$ values point to a shift of this bias towards exploitation, which may not always consist of `useful' exploitation, as will be discussed later in this section. In addition to quantifying the performance of the swarm, the tracking ability, $\Xi$, can also be used to quantify the effectiveness of the swarm's exploitative actions (Sect.~\ref{sec:metrics}); high levels of $\Xi$ serve as an indicator of a high level of effective exploitation. Note that in what follows, any optimum sought will always be with respect to maximizing $\Xi$.
\begin{figure}[htbp]
    \centering
    \includegraphics[width=\linewidth]{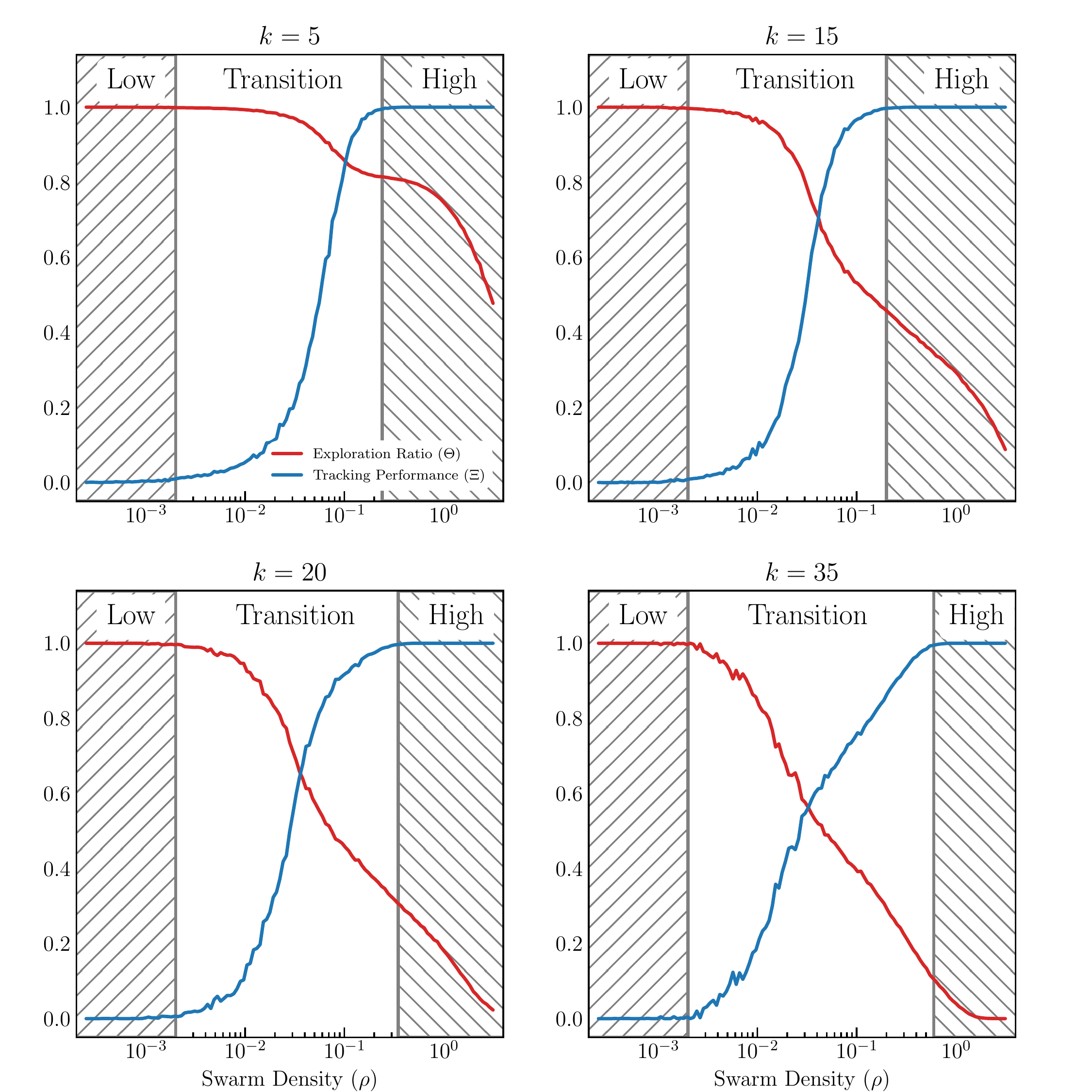}
    \caption{Tracking performance, $\Xi$, and exploration ratio, $\Theta$, of a swarm of $N=50$ agents using with levels of connectivity $k \in [5, 15, 25, 35]$, while tracking a non-evasive target traveling at {$\bv_{o, \text{max}} = 0.15$}. The low, transition, and high density phases have been marked out on the plots}
    \label{fig:score-explore}
\end{figure}

The existence of a trade-off between exploration and exploitation for collective tasks has previously been reported in different MAS~\citep{Oliveira2017, Kwa2022b}. However, a clear understanding of the underpinning of this trade-off is lacking. Here, we provide a quantitative analysis of this EED balance across the density spectrum. Starting with the low density phase identified in Fig.~\ref{fig:score-size}, we observe that the EED balance is fully skewed towards exploration regardless of the level of connectivity, $k$, (Fig.~\ref{fig:score-explore}). At such low swarm density, even heavily favoring the conduct of exploitative actions through high levels of $k$ is insufficient to yield minimal tracking of the target. In other words, the swarming strategy is plainly ineffective. As the system moves into the transition phase in Fig.~\ref{fig:score-explore}, the trade-off between exploration and exploitation becomes apparent. This is rooted in the fact that small inter-agent distances make it easier for individuals to flock together, in turn promoting the exploitation of any target information picked up by exploring agents. As $\Xi$ goes up, more resources become allocated to exploitation, causing the exploration ratio, $\Theta$, to go down in tandem. In this transition phase, the influence of the connectivity, $k$, is clearly marked. As reported in the previous section, the rise in $\Xi$ occurs at lower density for high $k$ systems compared to low $k$ ones. Specifically, when looking at the $k=35$ case in Fig.~\ref{fig:score-explore}, we notice a rapid decrease in $\Theta$ across the entire transition phase, almost to the point where no exploration is carried out at all. This should be compared to the case $k=5$, for which at the higher density end of the transition phase, high $\Xi$ values are achieved while still maintaining a relatively high level of exploration, $\Theta$. It is therefore interesting to notice that in that case, this trade-off is not associated with mutually exclusive actions; it is still possible to maintain high levels of exploration while carrying out similarly high levels of exploitation at intermediate to high densities ($8 \times 10^{-2} \lesssim \rho \lesssim 1 \times 10^{0}$). This is congruent with the current literature where it has been shown that systems are able to carry out both exploration and exploitation simultaneously~\citep{Kwa2022a}. Finally, when moving to the high density phase, which is exploration-limited, we are better able to understand why low $k$ systems outperform high $k$ ones. Indeed, with a low connectivity, maximum tracking performance is achieved while still allocating resources to exploration. This effective EED balance at $k=5$ therefore helps sustain high levels of system's vigilance, which is key to achieving the coveted high tracking scores. In comparison, with $k=35$, the rise in $\Xi$ across the transition phase, although starting earlier, is not as rapid as in the case $k=5$. For $\rho \gtrsim 10^{-1}$ (in the right half of the transition phase), the rise in $\Xi$ markedly slows down, thereby requiring fairly high swarm densities to achieve perfect tracking. This noticeable drop in performance is traced to the reduced system's vigilance given the rapid drop in $\Theta$. 

This rich and complex set of behaviors highlights an important fact: adaptivity in the MAS behavior is required when varying the swarm density. Indeed, at a given density, adaptivity allows one to seek the optimal EED balance that yields maximal tracking. Note that in the present search and tracking problem, the level of connectivity $k$ serves as our `adaptivity' lever. At this stage, the next question arising is how to identify the optimum EED balance (i.e., the value of $k$ here) at any given density $\rho$.

\begin{figure}[hbt!]
    \centering
        \begin{subfigure}[t]{0.475\linewidth}
            \centering
            \includegraphics[width=\linewidth]{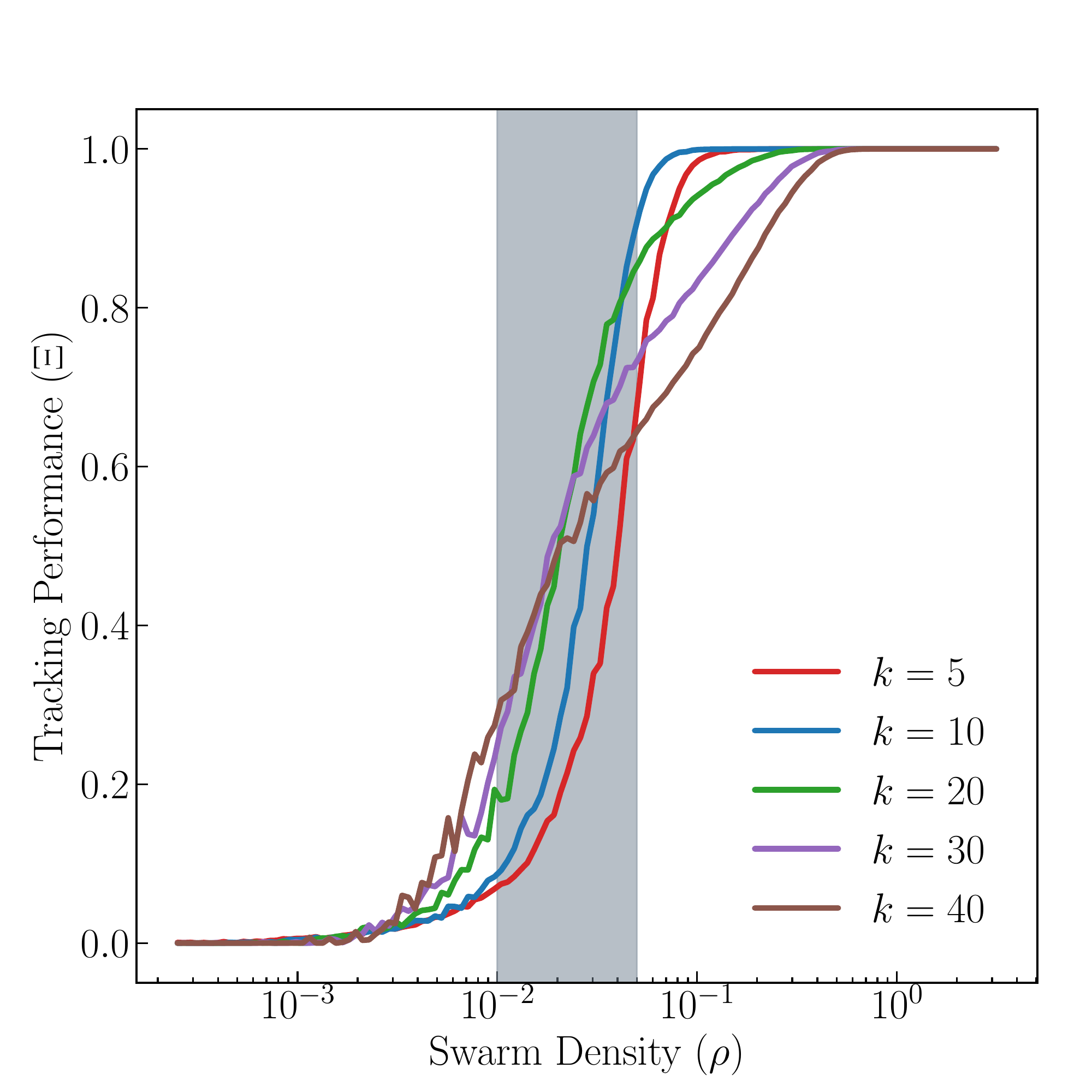}
            \caption{} 
            \label{fig:score-dens}
        \end{subfigure}
        \begin{subfigure}[t]{0.475\linewidth}
            \centering
            \includegraphics[width=\linewidth]{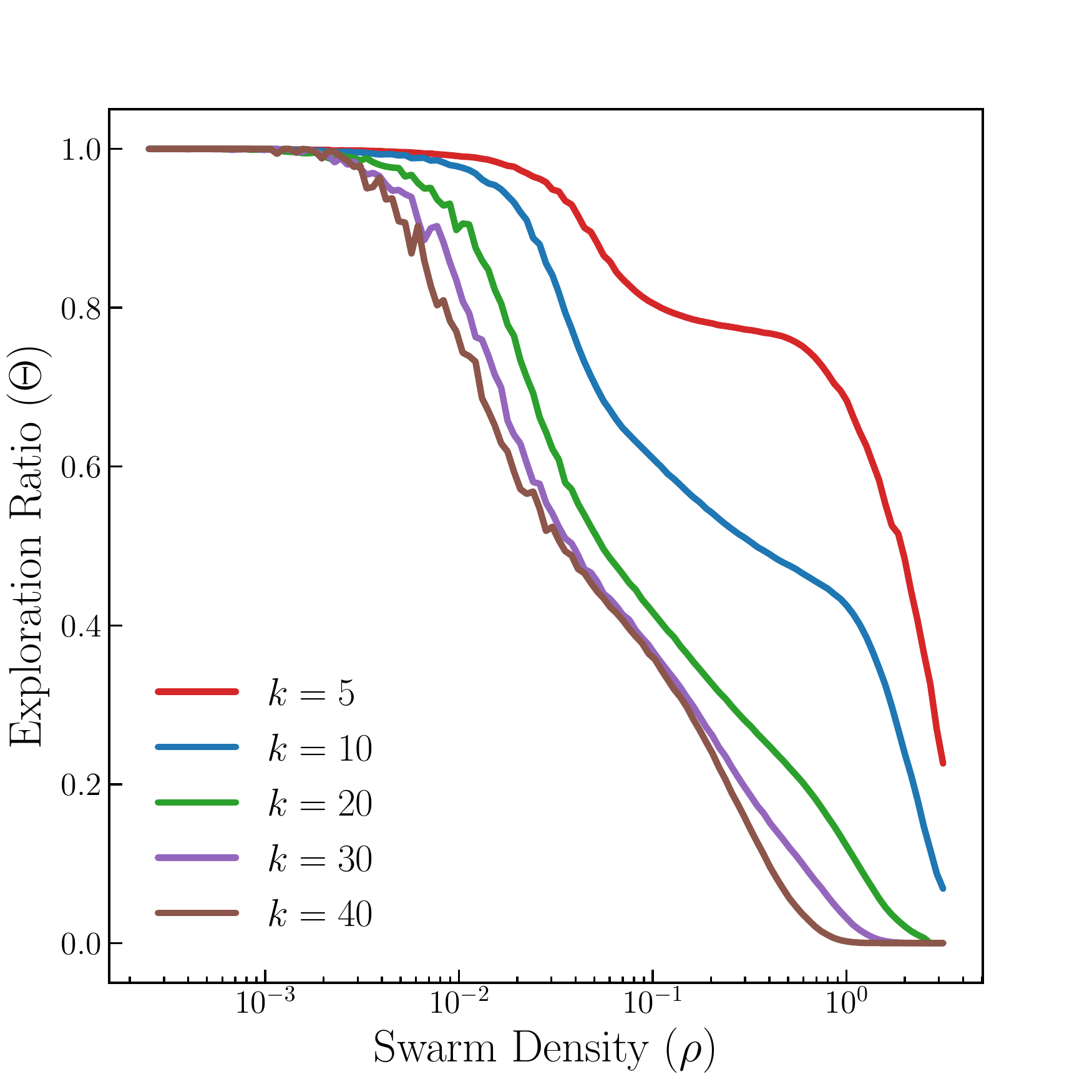} 
            \caption{} 
            \label{fig:explore-dens}
        \end{subfigure}
        \begin{subfigure}[t]{0.475\linewidth}
            \vspace{-2ex}
            \centering
            \includegraphics[width=\linewidth]{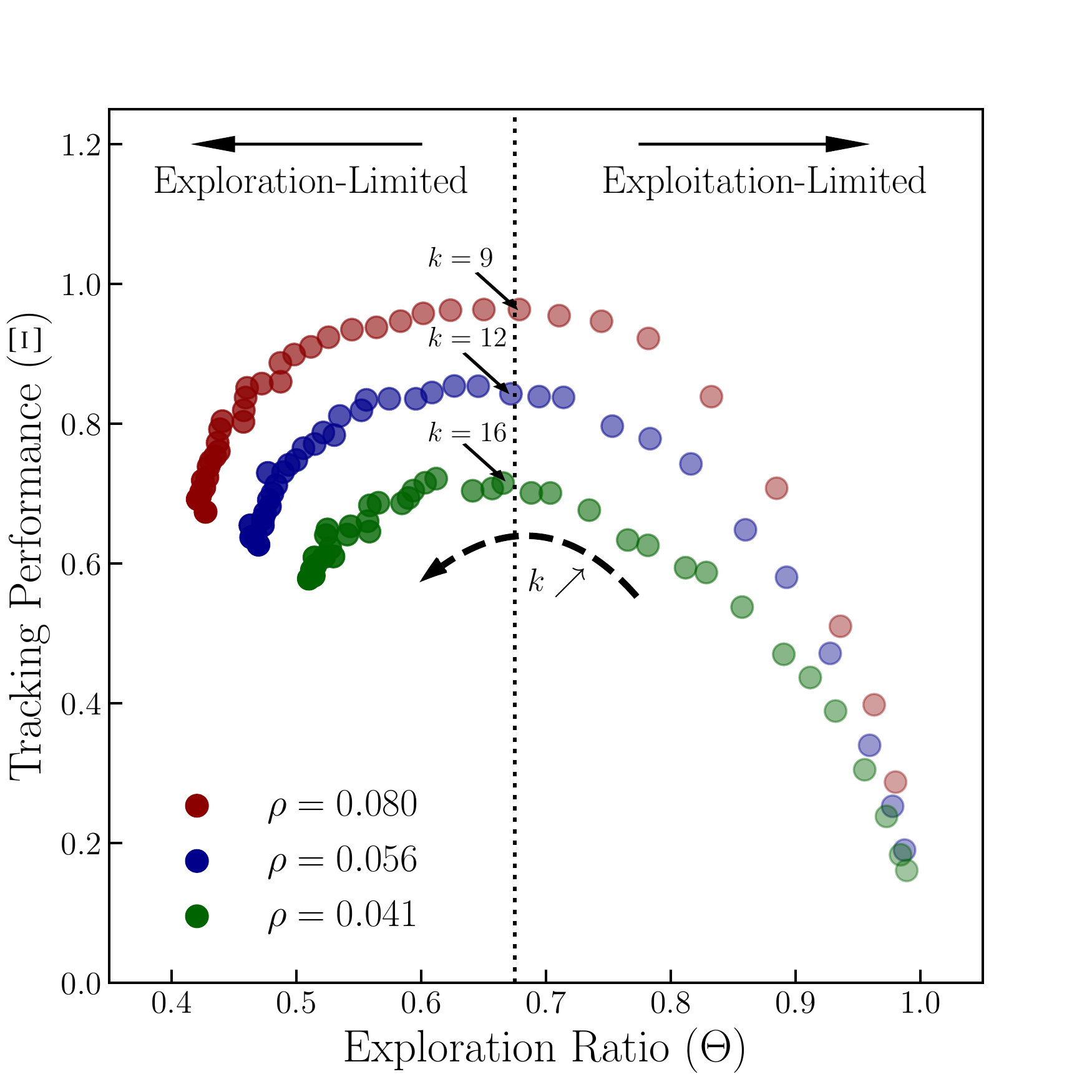}
            \caption{} 
            \label{fig:balance}
        \end{subfigure}
    \caption{A swarm of $N=50$ agents tracking a fast-moving non-evasive target traveling at $\bv_{o, \text{max}} = 0.15$. \ref{fig:score-dens}: Relationship between a swarm's tracking performance ($\Xi$) and the average swarm density ($\rho$) while operating at different levels of connectivity ($k$). The crossover region, where the optimum $k$ maximizes $\Xi$, is highlighted in gray. \ref{fig:explore-dens}: Relationship between a swarm's exploration ratio ($\Theta$) and the average swarm density while operating at different levels of connectivity. \ref{fig:balance}: Relationship between the system's tracking performance and exploration ratio. Lighter shaded points represent systems with lower $k$ while darker shaded points represent systems with higher $k$. The optimum level of exploration, $\Theta^*$, is represented by the dotted line with exploration-limited systems lying on the left of that line, while exploitation-limited systems lie on the right. All three densities considered fall within the transition region (Fig.~\ref{fig:score-size}).}
\end{figure}

Since there are different exploration-exploitation balances associated with different permutations of swarm densities and levels of connectivity, there exists an optimum level of connectivity for each density. This is illustrated in Fig.~\ref{fig:score-dens} (gray-shaded region) that reveals the existence of a crossover in performance within the transition region. Specifically, at the lower end of the transition phase, a higher level of tracking is achieved with a higher level of connectivity. However, as the density is gradually increased, the converse becomes true.

To gain better insights into these phenomena, we turn to the variations of the exploration ratio, $\Theta$, with $\rho$. Low degrees of connectivity, $k$, are associated with higher levels of exploration across the density spectrum (Fig.~\ref{fig:explore-dens}). Indeed, a reduced connectivity limits the spread of the target's positional information to a small subset of agents. Consequently, the agents beyond that subset are left to explore the surroundings. Since swarms with low $k$ tend to be exploitation-limited, they perform worse when operating at low $\rho$, where the conditions are known to hamper the conduct of exploitative activities. Conversely, high $k$ values promote the social transmission of information between agents. Thus, agents are better able to exploit this information collaboratively and cluster around the target. This higher level of exploitation allows high $k$ systems to outperform low $k$ systems starting from the lower end of the transition phase (prior to the crossover).

As expected, when increasing the density within the transition phase, the reduced inter-agent distances promote exploitation and the ensuing clustering of agents. Interestingly, at the higher end of the transition phase (past the crossover), low $k$ systems outperform high $k$ ones. Indeed, at these densities, such high $k$ systems exhibit over-exploitative actions, which can be qualified as `redundant' since excessive agent clustering around the target does not contribute to the tracking performance. As a consequence, these systems at these densities end up being exploration-limited.

The fact that a system's optimal level of connectivity changes with $\rho$ can be further explained using Fig.~\ref{fig:balance} that shows the presence of an optimum level of exploration, $\Theta^*$, at which tracking performance is maximized. As already mentioned, for a given swarm density, $\rho$, there exists an optimal level of connectivity, $k^*$, such that $\Xi(k^*, \rho)=\max_k(\Xi(\rho))$ (Fig.~\ref{fig:score-dens}). In turn, this gives us that $\Theta^*=\Theta(\max_k(\Xi))$. Although this optimum exploration level, $\Theta^*$, cannot easily be deduced from Fig.~\ref{fig:explore-dens}, it is readily apparent in Fig.~\ref{fig:balance} (see vertical dotted line). Interestingly, while determining the exact position of $\Theta^*$ requires further study, it appears to remain constant even as $\rho$ changes, which is connected to the changes in $k^*$ from one swarm density to the next. Furthermore, $\Theta^*$ clearly delineates between two regions corresponding to exploration-limited and exploitation-limited systems--- i.e., systems found to the left of the optimum are limited by their lack of exploration while systems on the right of the optimum have their performances limited by a lack of exploitation. This leads to the important conclusion that a system's level of connectivity---our `adaptivity' lever---must therefore be adjusted to manage and achieve an optimum balance between exploration and exploitation; a result previously suggested in several studies~\citep{Kwa2020a, Kwa2021, Talamali2021, Horsevad2022}.

\subsection{Effect of Swarm Size}
\label{sec:swarm-size}

In previous sections, we have reported compelling evidence that the concept of swarm density is key to the analysis of any MAS. A minimum density is necessary to trigger effective emergent collective actions. Furthermore, as the density increases and we enter the transition phase, the existence of an intricate balance between exploratory and exploitative actions reveals that these emergent properties are strongly dependent on the density. Nonetheless, one should not put aside unavoidable finite-size effects;
for instance, those that arise due to the limited number of agents employed within any MRS, usually constrained by financial and logistical challenges~\citep{Schroeder2019, Horsevad2022a}. As such, we must acknowledge the role that a system's size plays, and it appears therefore necessary to complement our density analysis by investigating the effects of swarm size.

\begin{figure}[hbtp]
    \centering
    \includegraphics[width=\linewidth]{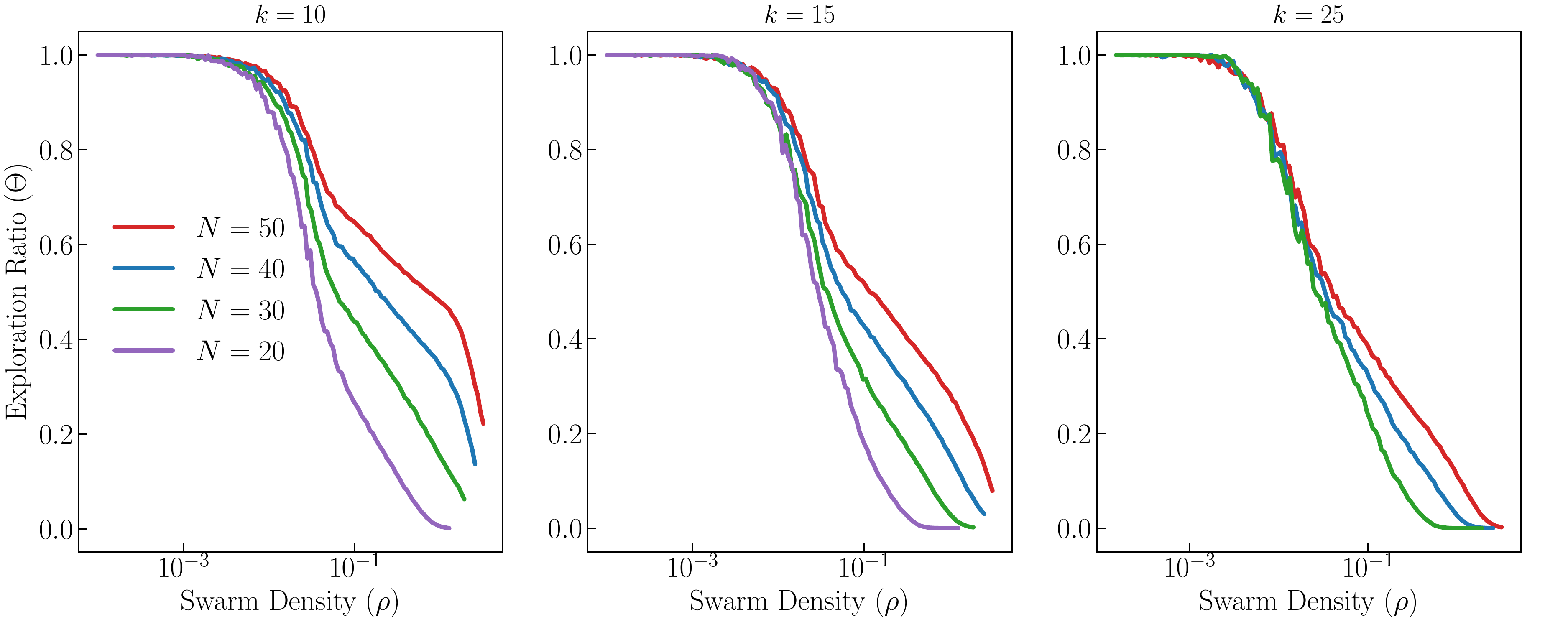}
    \caption{Exploration ratio of a swarm of $N=50$, $40$, $30$, and $20$ agents for three different levels of connectivity $k$ and varying swarm densities $\rho$ tracking a fast-moving non-evasive target traveling at $\bv_{o, \text{max}} = 0.15$.}
    \label{fig:explore_size}
\end{figure}

\begin{figure}[hbtp]
    \centering
    \includegraphics[width=0.5\linewidth]{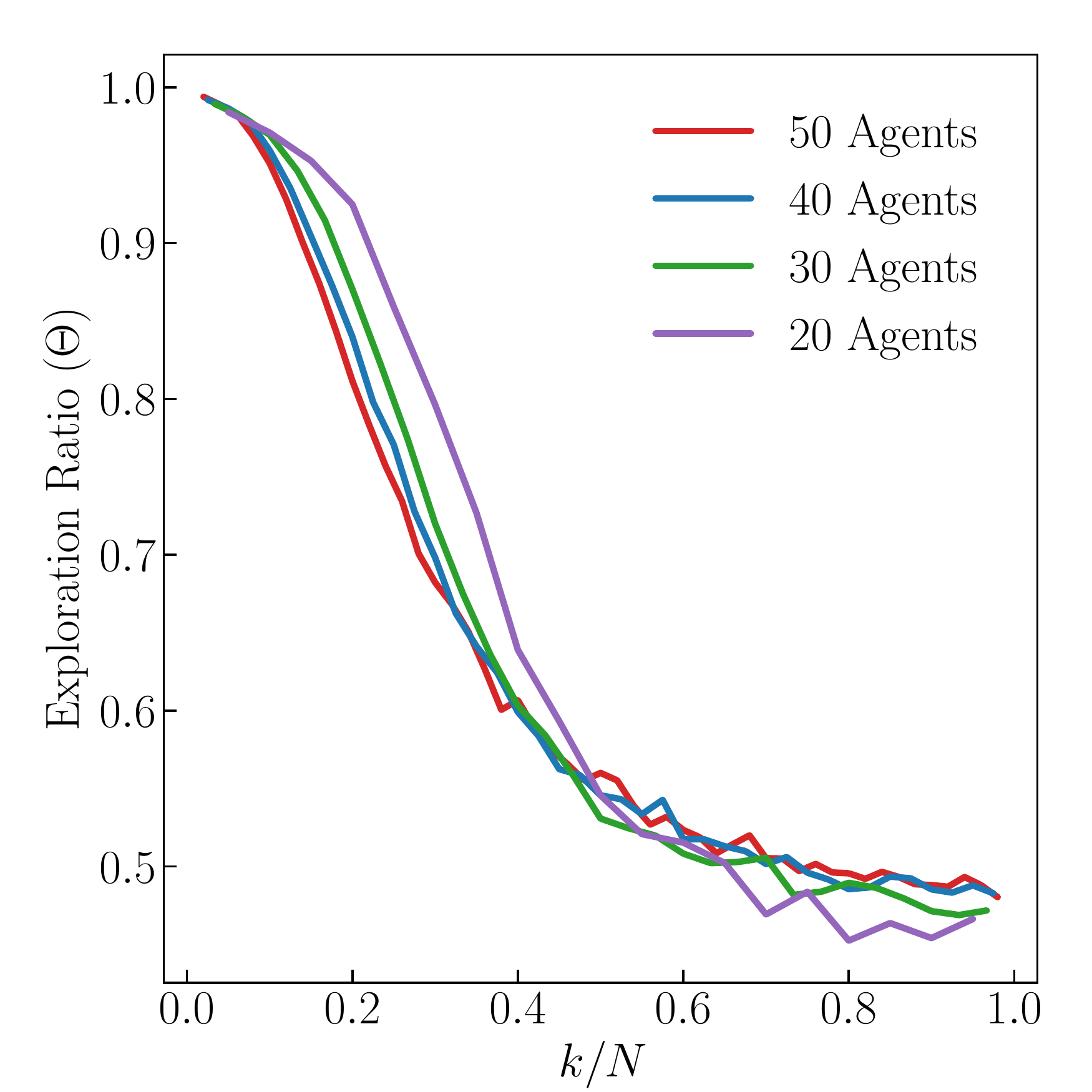}
    \caption{Exploration ratio of a swarm of $N=50$, $40$, $30$, and $20$ agents with varying levels of connectivity at a swarm density of $\rho=0.0444$, tracking a fast-moving non-evasive target traveling at $\bv_{o, \text{max}} = 0.15$.}
    \label{fig:explore_scale}
\end{figure}

As previously discussed, in natural animal collectives and computational optimization problems, increasing the number of agents within a system primarily serves to increase the level of exploration carried out. Figure~\ref{fig:explore_size} illustrates that this is also the case in the target tracking swarm. Especially as the swarm density increases and when operating at high levels of connectivity, the figure shows that large systems tend to maintain higher exploration ratios compared to small systems. This is because the level of exploration carried out by the swarm varies according to $k/N$, the number of neighbors an agent has as a ratio to the total size of the swarm, as shown in Fig.~\ref{fig:explore_scale}. These results suggest that exploration is a system-level action and therefore depends on the proportion of agents carrying out exploratory activity. When one agent transitions from exploration to exploitation, small swarms lose a larger amount of their exploration capacity compared to large swarms (10\% in a swarm comprised of 10 agents compared to 2\% in a swarm comprised of 50 agents). As such, when using similar levels of connectivity, large systems tend to have a larger proportion of agents that remain unengaged with the target, allowing them to sustain higher levels of exploration compared to systems comprised of fewer units.

\begin{figure}[hbtp]
    \centering
    \includegraphics[width=\linewidth]{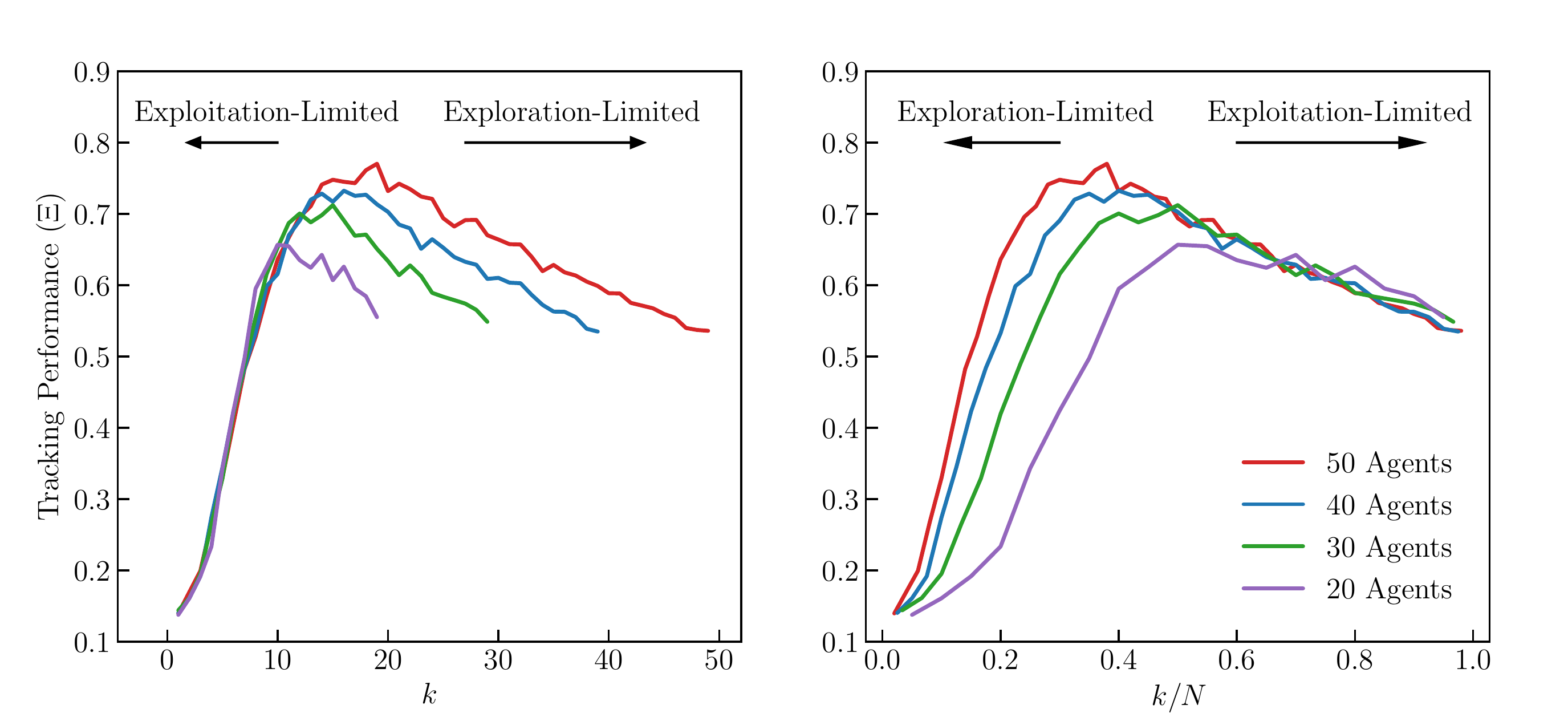}
    \caption{Tracking performance of a swarm of $N=50$, $40$, $30$, and $20$ agents with varying levels of connectivity at a swarm density of $\rho=0.0444$, tracking a fast-moving non-evasive target traveling at $\bv_{o, \text{max}} = 0.15$. Tracking performances are plot against $k$ (left) and $k/N$ (right).}
    \label{fig:track_scale}
\end{figure}

The system-level nature of exploration can also be observed when analyzing the tracking performances of differently sized systems. Figure~\ref{fig:track_scale} illustrates that there also exists an optimal level of connectivity ($k^*$) maximizing a system's tracking performance. At $k<k^*$, the system operates in the exploitation-limited phase. Conversely, at $k>k^*$, the system belongs to the exploration-limited phase. Therein, it can be seen that all systems experience similar losses in performance that are proportional to $k/N$, regardless of system size. This further highlights that the amount of exploration carried out by a system is effectively a global property.

In contrast, when operating in the exploitation-limited phase, it can be seen that the systems' performances are proportional to $k$, the absolute number of neighbors, regardless of the system size. This shows that exploitation is a local-level task that depends on the number of agents that are able to cluster around the target. When operating in the exploitation-limited phase, using the same level of connectivity results in the same number of agents being drawn to the target, thereby keeping the level of exploitation constant across systems with different numbers of agents.   

Besides the trends observed in exploration and exploitation, Fig.~\ref{fig:track_scale} also reveals that small swarms display lower maximum tracking performances, as well as a lower $k^*$, when compared to a large swarm. This can be traced to the lower capacity of small swarms to concurrently carry out exploratory and exploitative actions. As the level of connectivity of a small swarm rises, the system rapidly dedicates larger proportions of agents to the exploitation task by having them engage with the target. This in turn reduces the proportion of agents carrying out exploration, causing the system to move from the exploitation-limited to the exploration-limited phase more quickly than for larger swarms. 

\begin{figure}[hbtp]
    \centering
    \includegraphics[width=\linewidth]{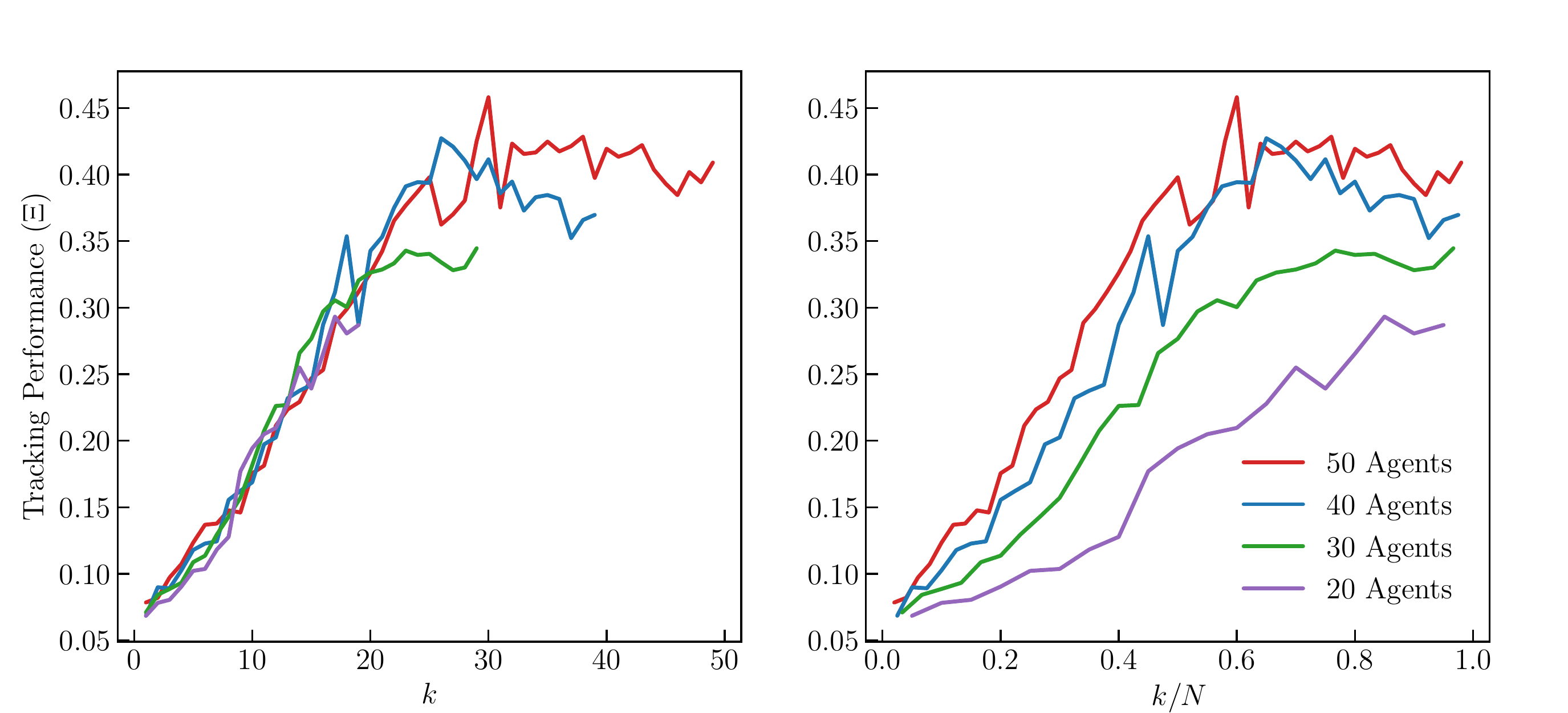}
    \caption{Tracking performance of a swarm of $N=50$, $40$, $30$, and $20$ agents with varying levels of connectivity at a swarm density of $\rho=0.02$, tracking a fast-moving non-evasive target traveling at $\bv_{o, \text{max}} = 0.15$. Tracking performances are plot against $k$ (left) and $k/N$ (right).}
    \label{fig:track_scale_low}
\end{figure}

The lack in exploitation capacity also explains the trends seen in Fig.~\ref{fig:track_scale_low} that shows the swarm's performance when it is operating closer to the low density phase. Here, it can be observed that the tracking performance still increases proportionally with $k$, further confirming that the task of target tracking, and hence exploitation, requires localized coordination. However, for the swarms comprised of 30 and 20 agents, tracking performance continues to rise when $k$ is increased and does not peak. This suggests that, due to the small number of agents and the large distances between them, the swarm is unable to fully exploit the target's positional information when it is found. As such, there is no optimal level of connectivity and no transition to the exploration-limited phase.

\section{Discussion}

When considering a system of $N$ robots that collaboratively performs a given task over a given domain area, one typically seeks the minimum number of units to accomplish that according to a specific metric. On one hand, with too few robots, the MRS inevitably fails to coordinate the actions of its individual units, thereby hindering the emergence of a robust collective response. On the other hand, with too many robots, unavoidable interference hampers the coordination of behaviors~\citep{Hamann2012,Hamann2018swarmrobotics}. This points to a range of values for $N$ that is neither too low, nor too high. More generally, this points to the existence of an interval in agents density over which, the system goes from poor (to nonexistent) collective action to a maximum in collective performance. 

For our collective task of target tracking, we have indeed identified such an interval, with a clear minimum density required to effectively trigger the necessary emergent coordination in the so-called exploitation-limited regime (Fig.~\ref{fig:score-dens}). At the other end of this interval, the upper bound in density corresponds to a permanent and sustained tracking of the target by the MRS, within the so-called exploration-limited regime. From the practitioners perspective, one should seek all possible ways to shift this interval towards the lower density values. This would amount to achieving better tracking performance with a smaller system size---i.e., at a reduced overall cost. In other words, advances in the design of the collective actions can help take a fuller advantage of swarm intelligence.

\begin{figure}[hbtp]
    \centering
    \includegraphics[width=0.65\linewidth]{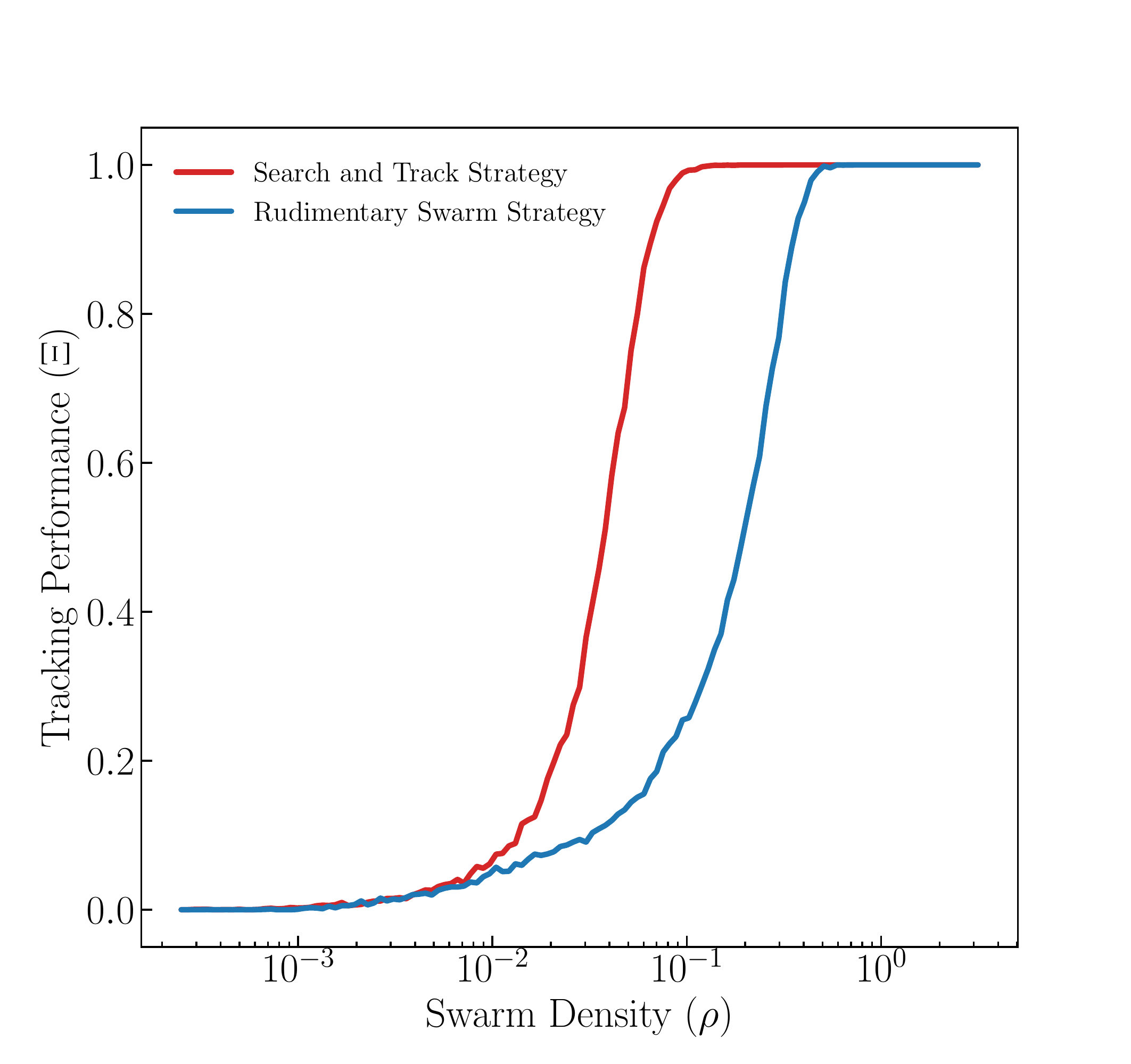}
    \caption{Comparison of the search and track strategy presented in Sect.~\ref{sec:search-and-track-strategy} and a rudimentary memoryless strategy that uses a constant inter-agent repulsion behavior. Both sets of simulations were carried out at a level of connectivity, $k=10$. The performance curve for the rudimentary strategy is shifted towards a higher density range compared to the baseline swarm strategy.}
    \label{fig:broken}
\end{figure}

To be specific, the swarm's flexibility is reflected by a particular exploration-exploitation balance, ultimately responsible for the observed variations of the tracking performance with the density. For the sake of this discussion, it is worth recalling the complete definition of flexibility given by~\cite{Dorigo2021}: ``The capacity to solve problems/perform tasks that depart from those chosen at design time." Needless to say that no absolute metric for flexibility has ever been proposed. We are therefore left to carry out relative comparisons of indirect flexibility measurements that have to be considered across a vast range of operating conditions. Indeed, for a given swarm strategy---i.e., a given set of local update rules---the effectiveness of the collective response of the system when the task/environment demands change is a proxy for flexibility. For our problem of target tracking, flexibility can be probed in many different ways: e.g., by changing the number of targets~\citep{Kwa2021, Kwa2022c}, the speed of those targets~\citep{Kwa2020a}, by adding obstacles~\citep{Sun2022}, etc. 

As a matter fact, at any given density, the only means to increase the collective tracking performance is to modify this exploration-exploitation balance by identifying potential adaptivity `levers'---``the ability to learn/change behavior to respond to new operating conditions"~\citep{Dorigo2021}. In the results presented in the previous sections, the only two adaptivity levers involved are: (1) the one associated with the memory introduced in the velocity attraction component (Appendix~\ref{sec:attraction-velocity-component}), and (2) the adaptive repulsion strategy (Appendix~\ref{sec:adaptive_repulsion}). This fact is clearly illustrated by Fig.~\ref{fig:broken}, where our search-and-track swarm strategy with $k=10$ (Sect.~\ref{sec:search-and-track-strategy}) is shown alongside a strategy, which is identical in all ways except that it is memoryless and has a fixed repulsion strength. As expected, both strategies are equally ineffective at low density. As the swarm density increases, the red tracking performance starts outstripping the blue one, then rapidly increases and peaks at one for a density almost ten times lower than the blue curve. The swarm strategy in red therefore exhibits much higher levels of flexibility than the blue one. For any density within the transition phase, the introduction of the inter-agent adaptive repulsion and memory (Sec.~\ref{sec:search-and-track-strategy}) taps into adaptivity that yields a more swarm-intelligent collective tracking of the target. Hence, in this context it appears that adaptivity promotes flexibility. It is worth adding that beyond memory and adaptive repulsion, other adaptivity levers exist and can be harnessed to further improve the effectiveness of the collective strategy. 

However, for this to be possible, it is key to have a clear understanding of the pivotal role played by the swarm density. As revealing as the previous analysis is, it does not stress enough the critical need to vary the density in order to: (i) identify in which phase the MRS is effectively operating---low density, transition phase or high density (Fig.~\ref{fig:score-size}), and (ii) characterize the actual exploration-exploitation balance and the associated level of flexibility. For instance, consider a practitioner designing and operating a given MRS with a given number of units $N$. Without varying the swarm density, this practitioner would not know whether the system is operating in the exploration-limited or exploitation-limited regime. Should the task require operations over a limited surface area, the density might be too high and swarm intelligence might not be necessary---it could induce interference effects hampering the collective dynamics. Arguably, the most interesting region is the transition phase, where adaptivity can be sought to boost swarm intelligence and deliver higher performances at the same density, or equivalently offer the same performance with a much lower number of units $N$. Beyond the need to identify the density interval corresponding to the transition phase, one also has to determine whether at that given density, the MRS behavior is exploration-limited or exploitation-limited. Failure to do so would significantly complicate the search for an effective adaptivity strategy. As was seen in Fig.~\ref{fig:score-dens}, in the lower range of densities within the transition phase, one can make use of the system's level of connectivity as another adaptivity lever and increase $k$. However, this strategy completely backfires in the exploration-limited regime for slightly higher densities, still in the transition phase. 

The previous analysis about the density dependence of robot swarms also helps explain some aspects of the simulation-reality gap for MRS. The latter has been acknowledged to be particularly wide owing to its typical characteristics as a complex system---i.e., the large number of interacting units responsible for the emergent behaviors~\citep{Francesca2016,Dorigo2021}. Indeed, challenges in adequately and accurately modeling and/or simulating robot swarms can lead to non-negligible differences in swarm density, which in turn would induce notable discrepancies between simulations and hardware experiments~\citep{Hamann2018swarmrobotics, Zhong2018a, Dorigo2021, Kwa2020a, Kwa2021}.

It is important to stress that the obtained results and analysis have been established for our problem of collective tracking. The question of how these results for this particular collective action can be generalized to other swarming behaviors naturally arises. At this point, it is worth recalling that few general results related to swarming behaviors effectively exist; there is no general scalability law, no robustness law, or flexibility law for that matter. The emergent nature of the collective action and the fact that swarms are complex systems can explain this dearth of general results, which is further compounded by the current lack of benchmark problems to quantitatively assess and compare the overall performance of a system with a given cooperative control algorithm~\citep{Kwa2022a}. Nonetheless, the key question about the influence of the swarm density on the emergent response of MRS remains applicable and valid for all robot swarms involved in a wide range of collective actions. As mentioned earlier, at extremely low density the swarm inevitably fails to self-organize. It is only by increasing the number of units---and thus the swarm density---can an emergent behavior materialize. At the other end of the density spectrum, a high swarm density means that so many units are available that swarm intelligence may not even be necessary to carry out the task at hand---too high a density may even hinder the collective organization. These well-known facts are unambiguous about the existence of a transition phase in which the system dynamics goes from lacking self-organization to exhibiting an emergent response. This transition phase has been observed in a number of collective behaviors including with Vicsek's model for self-propelled particles~\citep{Vicsek1995,Bouffanais2016}. Although the existence of this transition phase is not fundamentally new, its implications for the design of robot swarms performing any task---beyond just collective tracking---has been overlooked, in particular in relation with the sizing of the system. When characterizing the transition phase, some differences should naturally be expected across a wide range of tasks and problems. For instance, the adverse effects of interference at high density can effectively induce a drop in performance as has been recognized for foraging tasks with a central depository~\citep{Rosenfeld2006, Hamann2012}. Nevertheless, a systematic analysis of this transition phase for the class of collective tasks involving an exploration--exploitation balance~\citep{Kwa2022a} would contribute to further generalizing some of the results reported here. We therefore hope that this work will encourage other researchers to study the influence of swarm density across a wider range of platforms and tasks in swarm robotics.

We can conclude that an analysis of the density dependence for an MRS offers means of comparing and quantifying swarm intelligence when it is engaged in a particular task. This is a necessary step towards optimizing the collective performance, i.e., when seeking to boost swarm intelligence. From the machine learning perspective, these results and conclusions have far-reaching implications. Numerous strategies have been considered to establish the local control rules that lead to sought-after global collective actions: e.g., automatic design~\citep{Francesca2016,Ligot2022towards}, multi-agent reinforcement learning~\citep{Kouzehgar2020,Zhang2021multi}, deep reinforcement learning~\citep{Huttenrauch2019deep}, etc. However, in most of these approaches, the number of units (or equivalently the swarm density) is a parameter that is considered constant during the learning/optimization process. Based on the above analysis of the density dependence, one has to be cautious about the fact that if the learning/optimization process is simply carried out at a given density, it has the potential to yield dismal performance should the circumstances suddenly change. For instance, consider a learned behavior for a density in the upper range of the transition phase (say $\rho\simeq 10^{-1}$ in Fig.~\ref{fig:score-dens}), which should offer a relatively high tracking performance. Given what was found in Sect.~\ref{sec:results}, the learned behavior will most likely find ways to promote exploration, which limits the tracking at that density. Should the density suddenly drop due to the loss of units or an expansion of the domain area, it is highly probable that the learned behavior optimized for exploration will underperform given the increased need for exploitation. Similarly, the identified influence of swarm size in the exploration-limited regime (Sect.~\ref{sec:swarm-size}) should also be taken into consideration when resorting to machine learning approaches.

This does not imply that machine learning strategies should not be considered for this class of problems. Instead, what our analysis suggests is to use machine learning strategies that aim at optimizing adaptivity across the density spectrum. In conclusion, one should aim towards ``adaptivity learning" for MRS, while avoiding the learning of a specific set of behaviors for a given task and circumstances. To help convey the deep meaning of adaptivity learning, we can return once again to Fig.~\ref{fig:score-dens}, and learning adaptivity would amount to finding the optimal value of $k$ that yields the highest tracking performance for any value of the swarm density. As mentioned earlier, with our search and track strategy (Sect.~\ref{sec:search-and-track-strategy}), another potential adaptivity `lever'---beyond memory and the repulsive behavior---is the system's connectivity level, $k$. 
A caveat with this adaptivity learning approach is that it does not seem trivial to generalize to any swarm strategy. Nonetheless, the identification of all adaptivity levers associated with any swarm strategy is a first step in that direction.

Lastly, it can be argued that learning adaptivity is equivalent to dynamically optimizing the balance between exploitative and exploratory actions at the system level when circumstances and/or operating conditions change. We believe that learning adaptivity and the optimization of the exploration-exploitation balance will offer promising new opportunities and avenues for the community working on the design of effective collective swarm dynamics.

\backmatter





\section*{Declarations}

\subsection*{Ethics approval and consent to participate}
Not Applicable

\subsection*{Consent for publication}
Not Applicable

\subsection*{Availability of data and material}
The data used in this study can be found in the following GitHub repository \url{https://github.com/hianlee/swarm-density-tracking}.

\subsection*{Competing Interests}
H. L. Kwa is employed as a Research Engineer and receives a salary from Thales Solutions Asia.
All other authors have no relevant financial or non-financial interests to disclose.

\subsection*{Funding}
This work was supported by the Thales Solutions Asia under the Singapore Economic Development Board Industrial Postgraduate Programme (EDB IPP) and the Natural Sciences and Engineering Research Council of Canada (NSERC), under the grant \# RGPIN-2022-04064.

\subsection*{Author Contributions}
Conceptualization: R.B.; Methodology: H.L.K. \& R.B.; Development of Simulation and Data Processing Tools: H.L.K.; Conduct of Experiments: H.L.K.; Data Analysis: H.L.K., J.P. \& R.B.; Manuscript Preparation and Review: H.L.K., J.P. \& R.B.

\subsection*{Acknowledgements}
Not Applicable





\begin{appendices}

\section{Strategy Velocity Components}\label{ap:vel_comp}
The search and track strategy given in Sec.~\ref{sec:search-and-track-strategy} produces a velocity vector comprised of two parts: (1) the attraction velocity component, $\bv_{i,\text{att}}[t]$, and (2) the repulsion velocity component, $\bv_{i,\text{rep}}[t]$. These two components are then combined to give a final agent velocity using Eq.~\ref{eqn:movement}, which is restated here:

\begin{equation}
    \bv_i[t] = \bv_{i,\text{att}}[t] + \bv_{i,\text{rep}}[t].
\end{equation}
In this section, we state how the values for $\bv_{i,\text{att}}[t]$ and $\bv_{i,\text{rep}}[t]$ are obtained. This strategy was first presented in \cite{Kwa2022c}.

\subsection{Attraction Velocity Component}
\label{sec:attraction-velocity-component}
The attractive component is used to encourage agents to aggregate at a point of interest, $\bp[t]$, determined using Algorithm~\ref{alg:set_p}. At every time-step, each agent measures its local environment to look for a target. Should an agent detect a target, the agent will transition from an exploratory state into a tracking state, set $\bp[t]$ as the target's current location, and broadcast the location. Should a target not be detected, the agent will communicate with its $k$-nearest neighbors and attempt to track targets detected by its neighbors. In addition, each agent is endowed with a memory, $M$, of a duration of $t_\text{mem}$. Using this memory, each agent is able to keep track of the position and time at which a target was found. Each agent also receives a set of target positions and encounter times from its $k$-nearest neighbors. These received values are compared to an agent's own values and the most recent target position is used as a point of attraction, $\bp[t]$. At this point, should the agent still not have any knowledge of the target's location, $\bp[t]$ is set to the agent's own location, $x_i[t]$, essentially disabling the attractive component. Through the use of this update algorithm, agents are able to compare information that is directly sensed from the environment with information received from its neighbors and choose which set of information to exploit.

At this point, it is important to reemphasize that in this framework, the neighborhood of an agent is to be understood in the network sense. As such, an agent $i$ has as many neighbors as its degree, $k$. Also, since time-varying network topologies are considered, it should be noted the neighborhood of each agent evolves over the course of the task duration. Given this dynamic network topology, all agents independently set $\bp[t]$ using Algorithm~\ref{alg:set_p}.

\begin{algorithm}
\caption{: Point of Attraction Update Algorithm}
\label{alg:set_p}
    \begin{algorithmic}
    
    \State Initialise $M = t_\text{mem}$

    \If{Agent detects target}
        \State $\bp_{\text{self}} \gets$ Target's position
        \State $t_{\text{best}} \gets t$
    \EndIf
    
    \State Determine $\mathcal{N}_i = \{j \in [1, N]$ s.t. agent $j$ is a topological \textit{k}-nearest neighbor of agent $i$\}
    
    \State Get list of all neighbors' $\bp$ and $t_{\text{best}}$
    \State $\bp_{\text{neigh}} \gets \text{Most recent entry in all neighbors' } \bp$
    \State $t_{\text{neigh}} \gets \text{Most recent entry in all neighbors' } t_{\text{best}}$ 
    
    \If{$t_{\text{best}} + M < t$} 
        \State $\bp_{\text{self}} \gets \emptyset$
    \EndIf
    \If{$t_{\text{neigh}} + M < t$} 
        \State $\bp_{\text{neigh}} \gets \emptyset$
    \EndIf
    
    \If{$\bp_{\text{self}} = \emptyset \textbf{ and } \bp_{\text{neigh}} = \emptyset$}
        \State $\bp[t] \gets \bx_i[t]$  
    \ElsIf{$t_{\text{best}} > t_{\text{neigh}}$}
        \State $\bp[t] \gets \bp_{\text{self}}$ 
    \Else
        \State $\bp[t] \gets \bp_{\text{neigh}}$ 
    \EndIf
    \end{algorithmic}
\end{algorithm}

Using an agent's velocity in the previous time-step and its location in relation to $\bp[t]$, the attraction component can be calculated according to:

\begin{equation}
    \bv_{i, \text{att}}[t] = \omega \bv_i[t-1] + c r \big(\bp[t] - \bx_i[t]\big).
    \label{eqn:vel_update}
\end{equation}
This equation is similar to that used in the social-only PSO model proposed by \citet{Engelbrecht2010}, where $\omega$ is the velocity inertial weight, $c$ is the social weight, and $r$ is a number randomly drawn from the unit interval. In computational optimization, this is the main driver of a the system's exploitative behaviour. Here, it is used to drive the MRS towards the target. It should be noted that in the proposed strategy, unlike the social-only PSO model that uses an infinite memory length, agents here instead use a limited memory length. This is done to prevent agents from exploiting outdated target positional information.

\subsection{Adaptive Repulsion}
\label{sec:adaptive_repulsion}
The adaptive repulsion component is used to promote agent exploration of the search space and stop agents from aggregating within a small area, thereby preventing over-exploitation of target information. In addition, this behavior also offers an anti-collision measure as a direct byproduct of this mechanism. 

The inter-agent repulsion scheme adopted is based on the one used in the BoB environmental monitoring swarm developed by \citet{Vallegra2018, Zoss2018}. Using this behavior, an Agent $i$ with topological neighbors $j$ calculates its individual repulsion velocity as follows:

\begin{equation}
    \bv_{i, \text{rep}}[t] = - \sum_{j\in \mathcal{N}_i}\left( \frac{a_R[t]}{r_{ij}[t]}\right)^d \frac{{\mathbf{r}_{ij}[t]}}{r_{ij}[t]}, 
    \label{eqn:repulsion}
\end{equation}
where $\mathbf{r}_{ij}$ is the vector from Agent $i$ to Agent $j$ and $r_{ij} = \|\mathbf{r}_{ij}\|$. This inter-agent repulsion is controlled by two parameters: the repulsion strength $a_R$, affecting the agents' distance from each other at equilibrium, and the exponent $d$ in the pre-factor term $(a_R/r_{ij})$. In the work carried out, $d$ is fixed at $6$ given that this value has very moderate effects on the performance of the EED strategy. At large $(a_R/r_{ij})$ and $d$ values, the repulsion strength of the agents is approximately equal to the nearest-neighbor distance in equilibrium configuration~\citep{Vallegra2018, Coquet2021}. 

The key aspect of this inter-agent repulsion is an agent's ability to adjust its own repulsion strength, $a_R[t]$, based on its local environment and neighbourhood. To this end, the agent's exploratory state, $S_{i, \text{exp}}[t]$, is used. When an agent has no target information, it enters an exploratory state, i.e., $S_{i, \text{exp}}[t] = 1$, it increases its $a_R$ value until a maximum value is attained. Conversely, if the agent is in a tracking state, i.e., $S_{i, \text{exp}}[t] = 0$, the agent gradually reduces its $a_R$ value until a minimum value is reached. The adaptive repulsion behavior used to obtain the repulsion component is summarized in Algorithm~\ref{alg:adaptive_repulsion}.

\begin{algorithm}
    \caption{Adaptive Repulsion}
    \label{alg:adaptive_repulsion}
    \begin{algorithmic}
    \State Initialise $a_{R,\text{min}}$, $a_{R,\text{max}}$, $d=6$, $\delta_{\text{explore}}=0.1$, and $\delta_{\text{track}}=0.75$
    \While{System active}
    \If{$S_{i, \text{exp}}[t] = 0$} // Agent in tracking state. Reduce $a_R$.
        \If{$a_R > a_{R, \text{min}}$}
            \State $a_R \gets a_R - \delta_{\text{track}}$
        \EndIf
    \ElsIf{$a_R < a_{R, \text{max}}$} // Agent in exploratory state. Increase $a_R$.
            \State $a_R \gets a_R + \delta_{\text{explore}}$
    \EndIf
    \State Calculate $\bv_{i, \text{rep}}$ using Eq.~\ref{eqn:repulsion}
    \EndWhile
    \end{algorithmic}
\end{algorithm}

\section{Local Density}\label{ap:loc_dense}
This section shows the reasoning why Equation~\ref{eqn:loc-dens} was used in the calculation of the system's local swarm density. This is restated here for completeness:

\begin{equation}
    \rho_L = \frac{1}{NT}\sum^T_{t=1}\sum^N_{i=1}\frac{7}{\pi L_{i,t}^2}.
\end{equation}

\begin{figure}[hbtp]
    \centering
    \begin{subfigure}[t]{0.45\linewidth}
        \centering
        \includegraphics[width=\linewidth]{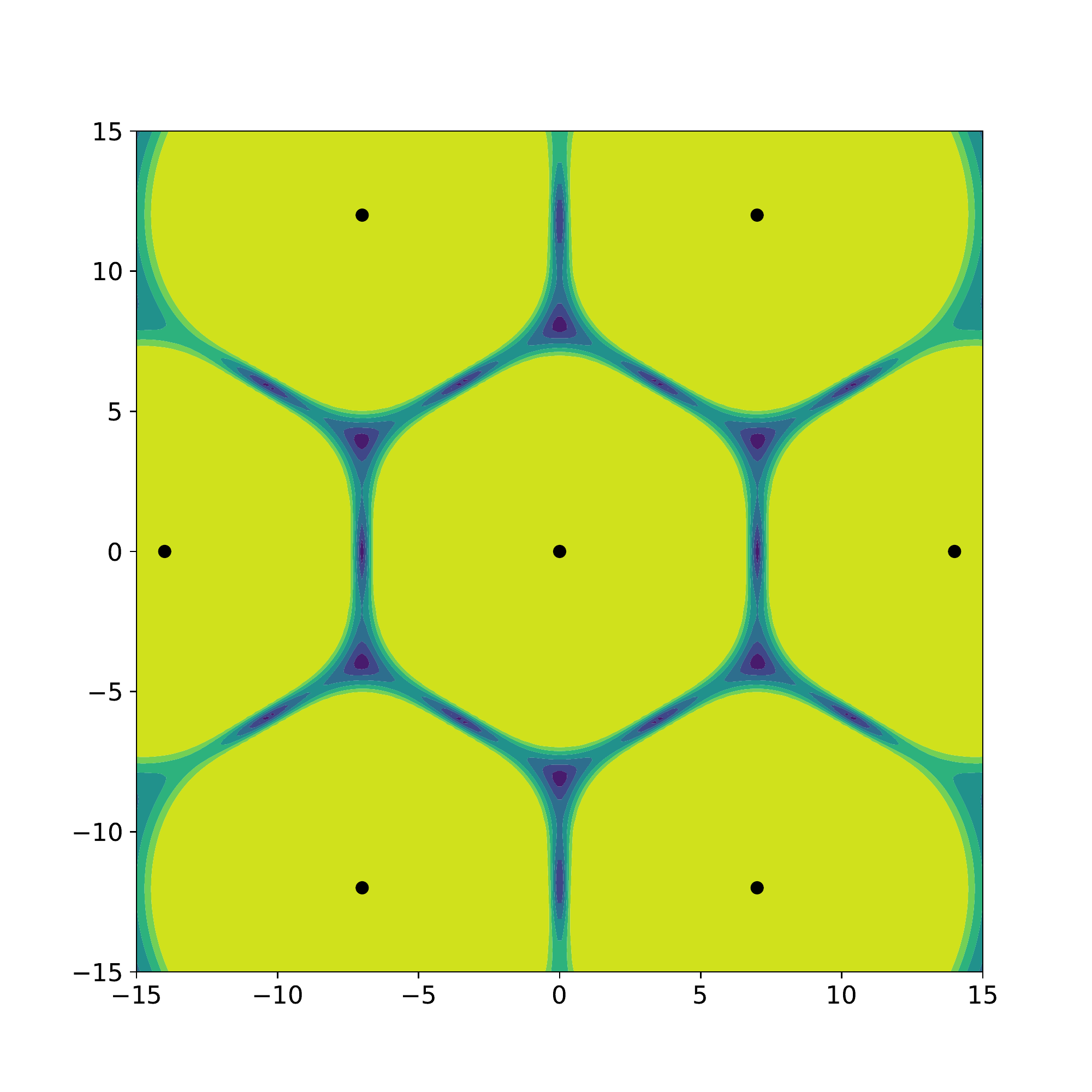}
        \caption{}
        \label{fig:eq-pos}
    \end{subfigure}
    \begin{subfigure}[t]{0.45\linewidth}
        \centering
        \includegraphics[width=\linewidth]{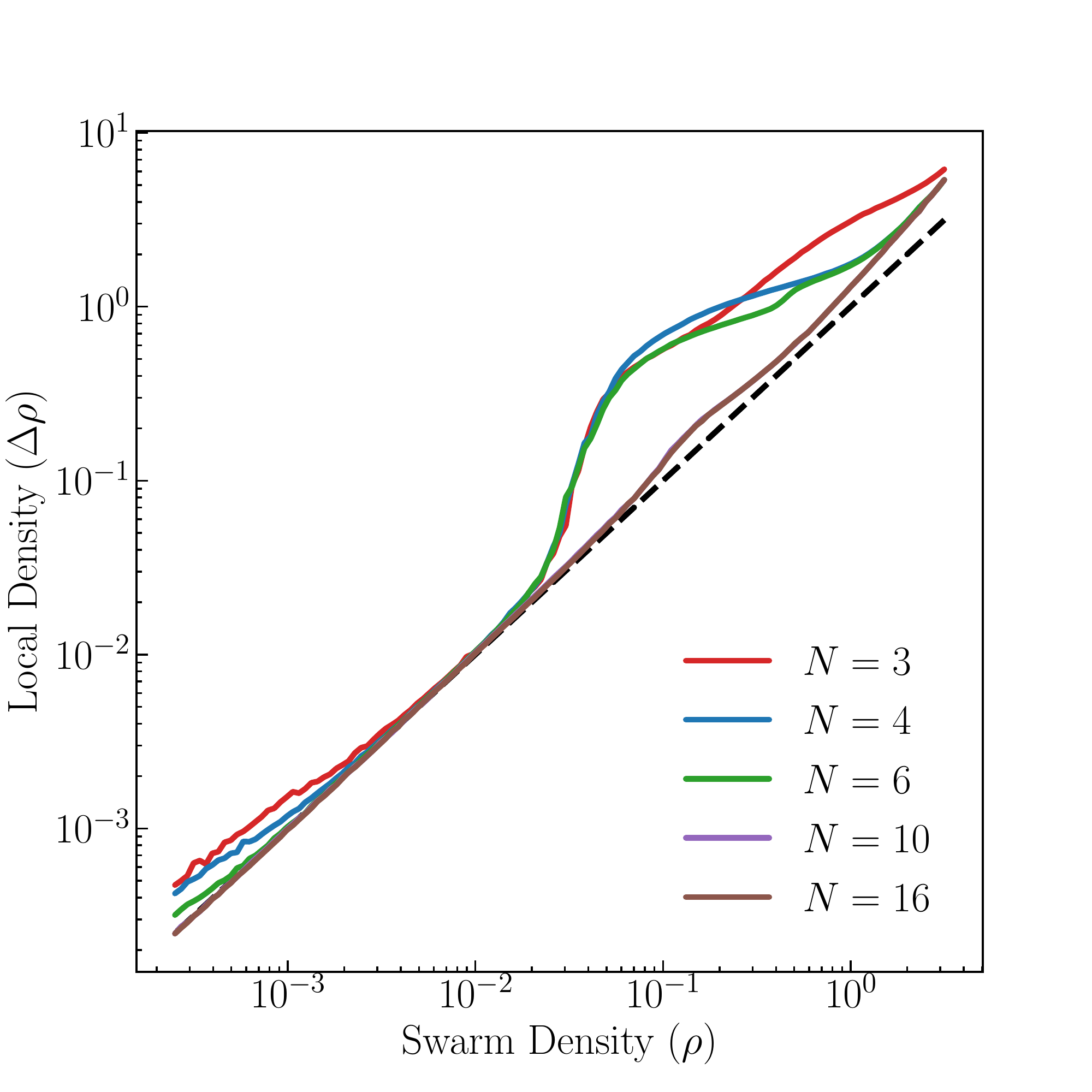}
        \caption{}
        \label{fig:loc-dense-metrics}
    \end{subfigure}
    \caption{\ref{fig:eq-pos}: Positioning of agents (black dots) in relation to each other and the repulsion fields generated by the individual agents while in their equilibrium positions. Areas in yellow represent areas of high repulsion potential while those in blue represent areas of low repulsion potential. Given these repulsion potential fields, agents tend to fall into a hexagonal packing pattern around each other. \ref{fig:loc-dense-metrics}: Local agent density calculated with different number of neighboring agents for a swarm comprised of 50 agents, connected using a $k=20$ communications network. The system is tracking a non-evasive target traveling at a maximum speed of $\bv_{o, \text{max}}=0.15$. The local agent density is compared against the global average swarm density (dashed line).}
    \label{fig:local-density}
\end{figure}

Due to the implemented inter-agent repulsion behavior, agents will tend to fall into a hexagonal packing pattern as seen in Fig.~\ref{fig:eq-pos}. As such, an individual agent will usually be surrounded by six other neighboring agents unless they are located at the edges of the system. By defining an $L_i$ as the average distance between an agent $i$ and its 6 nearest neighbors, it can be assumed that all 6 neighboring agents are located a distance of $L_i$ away for the purposes of calculation of an individual agent's local agent density. 

While a different number of agents can be used for this calculation, the same trends in the local agent density when varying the global average swarm density, as seen in Fig.~\ref{fig:loc-dense-metrics}. However, if less agents are used in this calculation, the initial divergence between the local agent density and the global average swarm density is accentuated. As such, an agent that finds itself in close proximity (relative to the size of the environment) to another agent while moving around the domain will return an artificially high local density. Conversely, if too many agents are used, the presence of such coincidental agent `clusters' is not reflected. As such, an intermediate number, six in this case, was chosen to be used for the local density calculations. 

\end{appendices}




\end{document}